\documentclass[pre,prd,twocolumn,superscriptaddress,showpacs,nofootinbib, amsmath, amssymb]{revtex4-1}

\usepackage{graphicx}
\usepackage{subfigure}
\usepackage{hyperref}
\usepackage{xcolor}
\usepackage[utf8]{inputenc}

\newcommand{\Gadget}{\textsc{GADGET-2}~}

\newcommand{\Rtr}{R_\mathrm{tr}}

\begin{document}

\title{Black Holes' Dark Dress: On the merger rate of a subdominant population of primordial black holes}



\author{Bradley J. Kavanagh}
\author{Daniele Gaggero}
\author{Gianfranco Bertone}
\affiliation{Gravitation Astroparticle Physics Amsterdam (GRAPPA),
Institute for Theoretical Physics Amsterdam and Delta Institute for Theoretical Physics,
University of Amsterdam, Science Park 904, 1090 GL Amsterdam, The Netherlands}

\begin{abstract}
The formation of astrophysical and primordial black holes influences the distribution of dark matter surrounding them. Black holes are thus expected to carry a dark matter `dress' whose properties depend on their formation mechanism and on the properties of the environment. Here we carry out a numerical and analytical study of the merger of dressed black holes, and show that the distribution of dark matter around them dramatically affects the dynamical evolution of the binaries. Although the final impact on the merger rate of primordial black holes is rather small with respect to the case of `naked' black holes, we argue that our analysis places the calculation of this rate on more solid ground, with LIGO-Virgo observations ruling out a dark matter fraction of $10^{-3}$ for primordial black holes of 100 solar masses, and it paves the way to more detailed analyses of environmental effects induced by dark matter on the gravitational wave emission of binary black holes. 
\end{abstract}

\maketitle

\section{Introduction}

In standard cosmology, most of the matter in the Universe is in the form of an elusive substance dubbed ``dark matter" (DM), which must be fundamentally different from the particles contained in the Standard Model of particle physics~\cite{Bertone:2016nfn,Bertone:2010zza}. In virtually all proposed formation scenarios, black holes form in environments characterized by a high dark matter density. Potentially large dark matter overdensities are expected to form around supermassive \cite{Gondolo:1999ef}, intermediate-mass \cite{Bertone:2005xz}, and so-called ``primordial'' \cite{2008ApJ...680..829R} black holes, as a consequence of their formation and evolution.

We focus here on primordial black holes (PBHs), i.e.~compact objects that may have formed in the early Universe from small-scale large-amplitude density fluctuations originated during inflation, or via a variety of other mechanisms (see e.g. \cite{Hawking:1971ei,Carr:1974nx,ChaplineNature1975}; for recent reviews, \cite{Green:2014faa,Sasaki:2018dmp}). The recent discovery of several gravitational wave signals from merger events of massive binary-black-hole (BBH) systems has prompted a renewed debate on the contribution of PBHs in the mass range $M_\mathrm{PBH}$ $\sim$ $1$ - $10^2$ M$_\odot$ to dark matter \cite{Bird:2016dcv,ClesseBellido2016}.

If one considers PBH pairs forming in virialized structures, the PBH binary merger rate is compatible with the one inferred by the LIGO and Virgo collaborations  ($\mathcal{R} \simeq$ 12 - 213 Gpc$^{-3}$ yr$^{-1}$ \cite{Abbott:2017vtc}) assuming that all of the DM is in the form of PBHs \cite{Bird:2016dcv}.  However, a significant number of PBH pairs decouple from the Hubble flow deep in the radiation era, therefore PBH pairs can copiously form in the early Universe as well \cite{Nakamura:1997sm,Ioka:1998nz}. A recent calculation of the associated merger rate at present time \cite{Sasaki:2016jop} -- extended and refined in \cite{Ali-Haimoud:2017rtz} -- provides a much larger estimate of the PBH merger rate. This can be translated into a bound on the fraction of DM in the form of PBHs, which is potentially much stronger than any other constraint in the same mass range (see \cite{PhysRevD.94.083504,Garcia-Bellido:2017xvr,Clesse:2017bsw,Gaggero:2016dpq} and references therein).

Here we show that the dark matter accumulated around PBHs  significantly modifies the dynamical evolution of PBHs binaries. 
In fact, unless PBHs contribute all of the DM in the Universe, they inevitably grow around them mini-halos of DM, whatever its fundamental nature is, in the early Universe ~\cite{Mack:2006gz,Ricotti:2007jk}. 
These DM ``dresses'' are expected to grow until the PBHs form binary systems that decouple from the Hubble flow, and to dramatically alter the evolution of the binaries due to {\it dynamical friction} \cite{Chandrasekhar1943a,Chandrasekhar1943b,Chandrasekhar1943c}.

While the PBHs orbit around each other, they interact with their respective DM mini-halos and induce slingshot effects on the DM particles,  losing energy and momentum in the process, and eventually heating up the DM halos. The effect of dynamical friction on BH binaries has been studied in the context of super-massive  binary systems at the center of galaxies, and is expected to make the binaries more compact and less eccentric. This can have a potentially significant effect on the merger time, and eventually on the merger rate of those objects at the present time (see, for instance, \cite{Begelman:1980vb,Quinlan:1996vp,2017PhRvD..96f3001G}).

In order to assess the impact of DM mini-halos on the orbits of PBH binaries, we perform N-body simulations to follow the dynamics of these systems, making use of the publicly available \Gadget code \cite{Springel:2005mi} as a gravity-only N-body solver. We follow the evolution of the PBH binary system from the time at which it decouples from the Hubble flow until the point at which most of the DM in the mini-halos has been ejected, and the semi-major axis and eccentricity of the system have stabilised. We thus self-consistently compute the merger times and merger rates of primordial BBH systems today, and compare it with the one inferred by the LIGO and Virgo collaborations. We also present a simple analytical model that captures the main aspects of the numerical calculations, and offers useful insights on the physics of the problem. 

The paper is organized as follows. In Sec.~\ref{sec:BBHproperties} we discuss the primordial BBH parameter space, following the formalism detailed in Ref.~\cite{Ali-Haimoud:2017rtz} and including the evolution of DM mini-halos following Refs.~\cite{Mack:2006gz,Ricotti:2007jk}; in Sec.~\ref{sec:simulations} we present the setup and results of our numerical simulations and our procedure to remap the BBH parameter space under the effects of dynamical friction. In Sec.~\ref{sec:results} we present our estimate of the merger rate that takes these effects into account; the resulting bound on the fraction of DM in the form of PBHs is shown in Fig.~\ref{fig:LIGO_limit}. Finally, we discuss these results and possible caveats in Sec.~\ref{sec:discussion}, followed by our conclusions in Sec.~\ref{sec:conclusions}. Code and results associated with this work can be found \href{https://github.com/bradkav/BlackHolesDarkDress}{here} \cite{DarkDressCode}.

\section{Formation and properties of PBH binaries formed in the early Universe}
\label{sec:BBHproperties}

\subsection{Properties of PBH binaries}
\label{sec:properties}

\begin{figure}[t]
\centering
   \includegraphics[width=0.95\linewidth,]{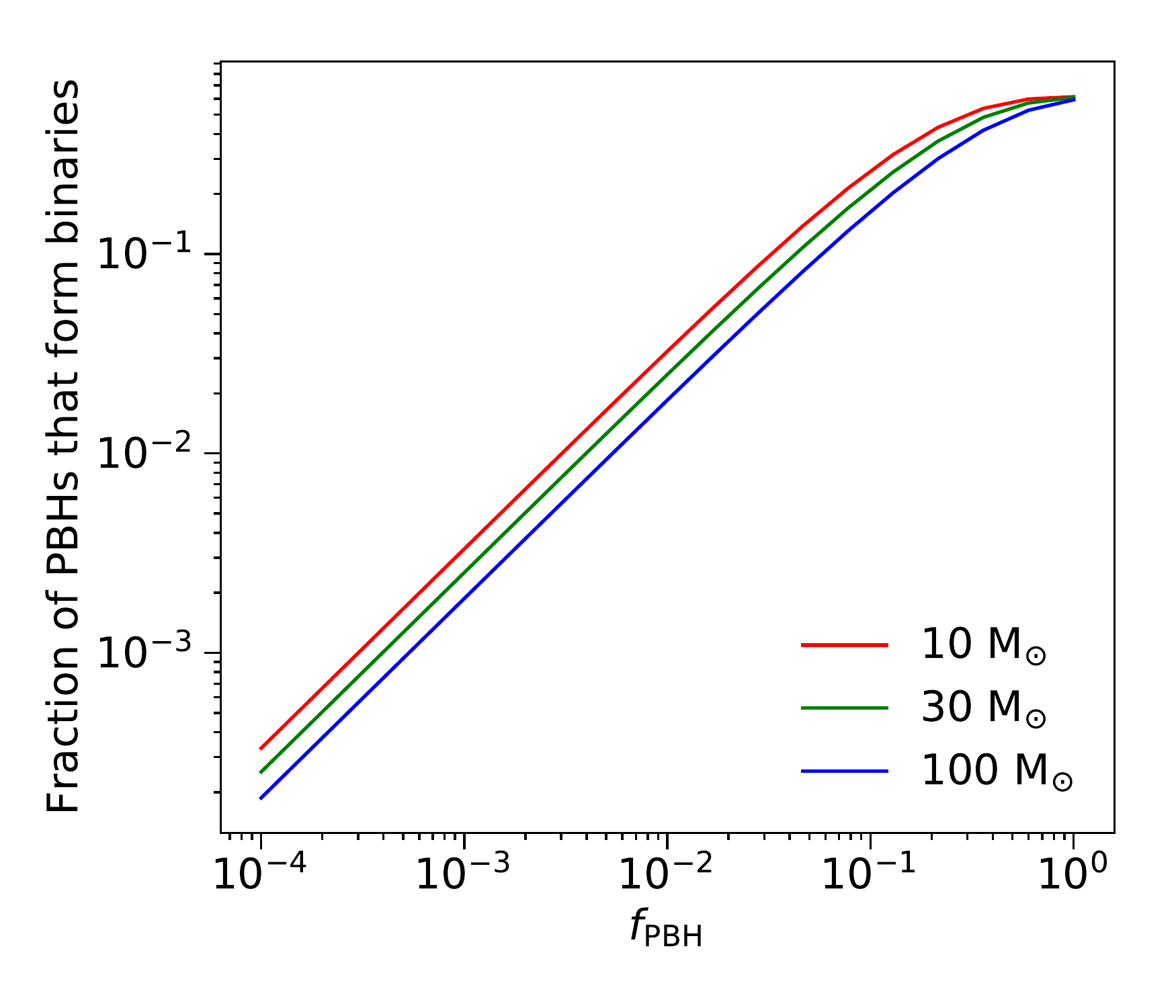}
   \caption{{\bf Fraction of PBHs that belong to some binary system formed in the early Universe}. This quantity is plotted as a function of the fraction of DM in PBHs  (for different values of the PBH mass). As mentioned in the text, if PBHs make all the DM, most of them belong to pairs that have a chance to decouple from the Hubble flow before matter-radiation equality and form a binary system.}
   \label{fig:fractionPBHplot}
\end{figure}

If PBHs make up all the DM, most of the pairs decouple from the Hubble flow before matter-radiation equality and form bound systems (see Fig. \ref{fig:fractionPBHplot}); if they only contribute a dark matter fraction $f_{\rm PBH} \ll 1$ only rare pairs with small separations form binaries. 
It is possible to determine the probability distribution for these systems in a two-dimensional parameter space where the two independent variables are the semi-major axis $a$ of the binary orbit and the dimensionless angular momentum, defined as 
\begin{equation}
j \equiv \frac{\ell}{\sqrt{2 G_N M_\mathrm{PBH} a}} = \sqrt{1 - e^2} \,,
\end{equation}
where $\ell$ is the angular momentum per unit reduced mass and $e$ is the eccentricity. 

Following the notation and the approach described in Ref.~\cite{Ali-Haimoud:2017rtz} (see also \cite{Chen:2018czv}), it is convenient to define the dimensionless variable $X$ as follows:
\begin{equation}
X \equiv \left(\frac{x}{\bar{x}}\right)^3 \,,
\label{eq:bigX}
\end{equation}
where $x$ is the comoving seperation of the PBH pair and
\begin{equation}
\bar{x} \equiv \left( \frac{3 M_{\rm PBH}}{4 \pi f_{\rm PBH} \, \rho_\mathrm{eq}}  \right)^{1/3}\,,
\end{equation}
is the mean (comoving) separation between two PBHs, in terms of the PBH mass $M_{\rm PBH}$, the density at matter-radiation equality $\rho_{eq}$, and the fraction $f_{\rm PBH}$ of DM in PBHs. Under the assumption that PBHs are uniformly distributed\footnote{The effect of clustering has recently been discussed in \cite{Ballesteros:2018swv}, and is found to be negligible for narrow mass functions, and potentially relevant for broader distributions.}, the differential probability distribution with respect to $X$ is simply
\begin{equation}
\frac{\partial P}{\partial X} \,=\, e^{- X}\,.
\end{equation}.

The angular momentum distribution is more tricky, as it requires us to model the tidal field the binaries are immersed in. 
A first estimate was performed in \cite{Nakamura:1997sm}, considering only the torque exerted by the tidal force caused by the PBH which is closest to the binary. 
A more refined treatment, accounting for the tidal torquing exerted by {\it all other PBHs} surrounding the binary itself, was presented later in \cite{Ali-Haimoud:2017rtz}. 

In the current work we adopt the latter prescription. It is useful to write explicitly the full PDF in terms of the variables $a$ and $j$ we are mostly interested in:
\begin{equation}
\label{eq:Paj}
P (a, j) |_{f, M_{\rm PBH}} \, = \, \frac{\partial X}{\partial a}\, \exp{\left(- \frac{x(a)^3}{\bar{x}^3}\right) } \, P(j)\,,
\end{equation}
with
\begin{equation}
\label{eq:measure}
\frac{\partial X}{\partial a} \,=\, \frac{\partial X}{\partial x} \frac{\partial x(a)}{\partial a} \,=\, \frac{3}{4 a^{1/4}} \left(\frac{f_{\rm PBH}}{\alpha \bar{x}}\right)^{3/4} \,,
\end{equation}
where:
\begin{itemize}
\item The relation between the decoupled binary semi-major axis $a$ and initial comoving separation $x$ of the PBH pair can be computed numerically \cite{Ali-Haimoud:2017rtz} by solving the equation of motion of two point sources subject to gravitational pull and Hubble flow at the same time:
\begin{equation}
\frac{{\rm d}^2 r}{{\rm d}t^2} = - \frac{2 G_N M_{\rm PBH}}{r^2} \frac{r}{|r|} + (\dot{H}(t) + H(t)^2) \, r\,,
\label{eq:Motion1}
\end{equation}
where $H(t)$ is the Hubble constant.

The solution clearly shows a turnaround of $r(t)$ followed by an oscillatory regime, which proceeds undisturbed by the Hubble flow; the relation between the semi-major axis $a$ of the newly formed binary and the initial PBH separation $x$ is then:
\begin{equation}
x(a) \simeq \left(\frac{3\,  a\, M_{\rm PBH}}{4 \pi \,\alpha \,\rho_{eq}} \right)^{1/4}
\label{eq:x_of_a}
\end{equation}
with $\alpha \simeq 0.1$ \cite{Ioka:1998nz,Ali-Haimoud:2017rtz}.

\item The  $j$ probability distribution is also estimated in the same paper, and can be written as follows:
\begin{equation}
\label{eq:Pj}
j \, P(j)|_{f, M_{\rm PBH}} \,=\,  \frac{y(j)^2}{\left(\, 1 \,+\, y(j)^2 \,\right)^{3/2}}\,,
\end{equation}
where 
\begin{equation}
y(j) \equiv \frac{j}{0.5\, (1+\sigma_{eq}^2/f^2)^{1/2}\,  X}.
\end{equation}
In the above expression, the contribution from large-scale Gaussian density perturbations, characterized by a variance $\sigma_{eq} \approx 0.005$ at matter-radiation equality, is taken into account.

\end{itemize}
With these prescriptions, the integral of the PDF over the full $(a, j)$ parameter space provides the fraction of PBHs that form a decoupled binary system in the early Universe, as shown in Fig.~\ref{fig:fractionPBHplot} for different values of the PBH mass and DM fraction in PBHs.

\begin{figure}[t]
\centering
   \includegraphics[width=\linewidth,]{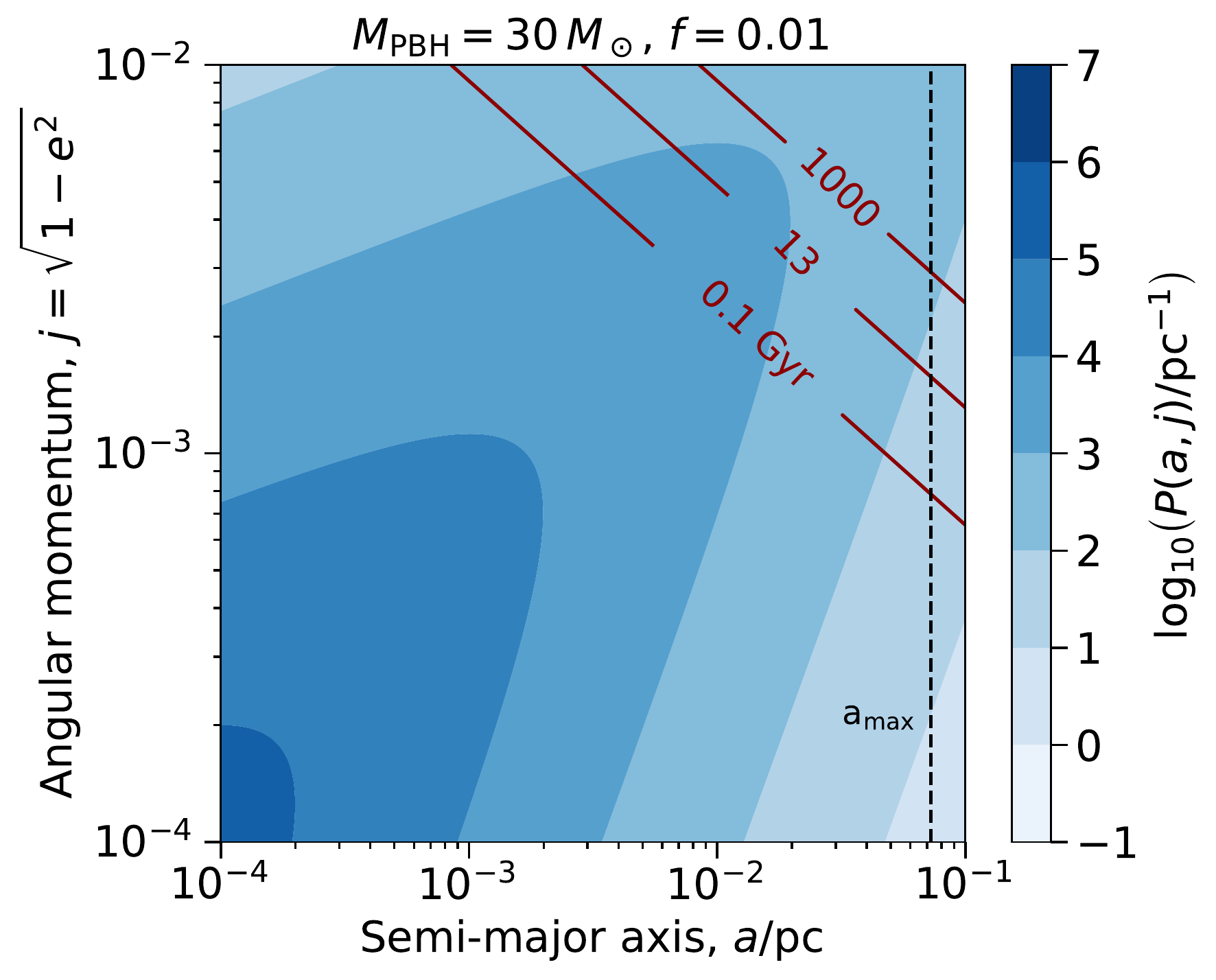}
   \caption{{\bf Probability distribution of PBH binaries that decouple in the early Universe}. The PDF, derived in \cite{Ali-Haimoud:2017rtz}, is given by Eq. \ref{eq:Paj}. We plot it as a function of the semi-major axis $a$ and dimensionless angular momentum $j = \sqrt{1-e^2}$. The red solid lines show contours of constant merger time (in Gyr).}
   \label{fig:ProbabilityDistribution}
\end{figure}

The full PDF $P(a,j)$ is displayed in Fig.~\ref{fig:ProbabilityDistribution}. In the same figure we also show the contours referring to the expected merger time of the binary due to the emission of gravitational radiation, which is given by \cite{Peters:1964zz}:
\begin{equation}
\label{eq:tmerge}
t_\mathrm{merge} \,=\, \frac{3 \, c^5}{170 \, G_N^3} \, \frac{a^4 j^7}{M_{\rm PBH}^3}\,.
\end{equation}
We remark that either a very small semi-major axis or an extreme eccentricity is required to get a merger time comparable with the age of the Universe ($t_\mathrm{univ} \sim 13.7 \,\,\mathrm{Gyr}$): wider, more circular binaries tend to merge on much longer timescales.

\subsection{Accretion of dark matter mini-halos before binary decoupling}
\label{sec:decoupling}
Let us now add another relevant piece of information to our model. 

Given the PDF described above, the authors of \cite{Ali-Haimoud:2017rtz} derived the merger rate at present time, and found that it would exceed the one observed by the LIGO and Virgo collaborations. Thus, PBHs can only be a small fraction of the DM in the Universe.

Motivated by these results, we consider a scenario characterized by a sub-dominant population of PBHs, immersed in a high-density DM-dominated environment, rapidly expanding and diluting. 
In this context, the relevant effect we want to model is the progressive growth of a DM mini-halo around each PBH, governed by the competition between the gravitational pull of the PBH and the expanding Hubble flow.

The accretion of the DM halo deep in the radiation era can be computed numerically~\cite{Mack:2006gz,Ricotti:2007jk}
by solving the following equation (similar to Eq. \ref{eq:Motion1}), describing radial infall of matter in an expanding universe:
\begin{equation}
\frac{{\rm d}^2 r}{{\rm d}t^2} = - \frac{G M_{\rm PBH}}{r^2} + (\dot{H} + H^2) r \, ,
\label{eq:Motion2}
\end{equation}
where $H(t) = 1/(2t)$. Evolving the above equation for each shell, starting from very high redshift with the initial conditions $r=r_i$ and $\dot{r} = H_i r_i = r_i/(2t_i)$, one finds that the PBH can accrete a DM halo with $M_\mathrm{halo}^\mathrm{eq} = M_\mathrm{PBH}$ at the end of the radiation era ($z = z_\mathrm{eq}$). 

The density profile of such a halo was first determined analytically in \cite{Bertschinger:1985pd} as a power law 
\begin{equation}
\label{eq:rhoDM}
\rho(r) \propto r^{-3/2}.
\end{equation}
We note that the same dependence on $r$ has been obtained in recent, realistic numerical simulations \cite{Delos:2017thv} that follow the evolution of ultra-compact mini halos (UCMHs)\footnote{Such halos can form out of small-scale large-amplitude density fluctuations that are too small to form PBHs, but still large enough to originate collapsed structures. The $\rho(r) \propto r^{-3/2}$ profile can develop if the UCMHs originate from a pronounced spike in the power spectrum at some given reference scale.}. There is however evidence that UCMHs may grow much shallower profiles or indeed that the profiles may be even steeper around isolated over-densities such as PBHs \cite{Gosenca:2017ybi}. In this work, we fix the density profile to that of  Eq.~\eqref{eq:rhoDM}, which is widely considered as a reference in the current literature.

Given the self-similarity of the profile, with no intrinsic length scale, it is useful to define a sharp cutoff at some {\it truncation radius}, which can be defined either as the turnaround radius, or the radius of the ``sphere of influence'' centered on the PBH and characterized by a DM density larger than the (rapidly declining) background density. 
According to \cite{Mack:2006gz,Ricotti:2007jk}, both definitions provide the same result:
\begin{equation}
R_\mathrm{tr} (z) \,=\, 0.0063 \, \left( \frac{M_\mathrm{halo}^\mathrm{eq}}{M_{\odot}} \right)^{1/3} \, \left( \frac{1 + z_\mathrm{eq}}{1 + z} \right) {\rm pc}\,,
\label{eq:r_tr}
\end{equation}
with respect to the redshift and the PBH mass. In the above expression, $_\mathrm{eq}$ refers to the quantities evaluated at matter-radiation equality (i.e.~at $z_\mathrm{eq} \simeq 3375$).

Hence, the halo mass accreted at a generic redshift $z$ can be written in terms of the truncation radius as follows:\footnote{We note that Eqs.~\eqref{eq:r_tr} and \eqref{eq:Mhalo} do not strictly apply \textit{before} matter-radiation equality. With this caveat, we apply them here. However, we also note that we will be mostly interested in binaries decoupling shortly before matter-radiation equality, where the true halo mass and size are unlikely to deviate by much from Eqs.~\eqref{eq:r_tr} and \eqref{eq:Mhalo}.}
\begin{equation}
\label{eq:Mhalo}
M_\mathrm{halo} (z) = \left(\frac{R_\mathrm{tr} (z)}{R_\mathrm{eq}}\right)^{3/2} M_\mathrm{PBH}\,.
\end{equation}

\begin{figure}[t]
\centering
      \includegraphics[width=0.5\textwidth,]{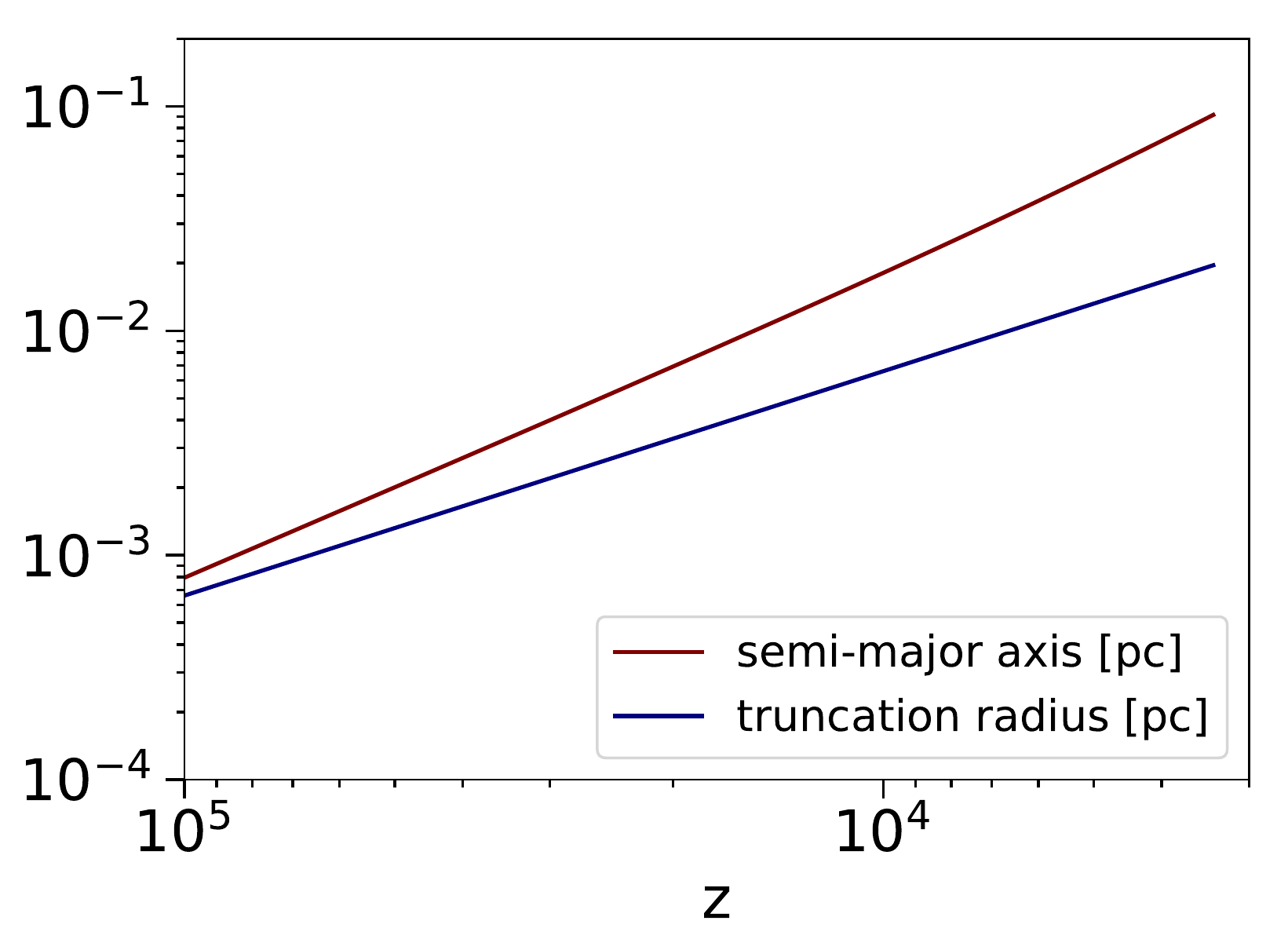}
      \caption{{\bf Truncation radius and semi-major axis of the decoupled binaries}. Both quantities are plotted as a function of the decoupling redshift (see Eq. \ref{eq:r_tr} and \ref{eq:z_of_a}). As mentioned in the text, we notice that more compact binaries decouple earlier (at larger redshifts). The plot shows that the DM halos never overlap in the redshift range considered here, because the truncation radius is always smaller than the semi-major axis of the binary system.} 
      \label{fig:r_tr_and_a}
\end{figure}

\begin{figure}[t]
      \centering
      \includegraphics[width=0.5\textwidth,]{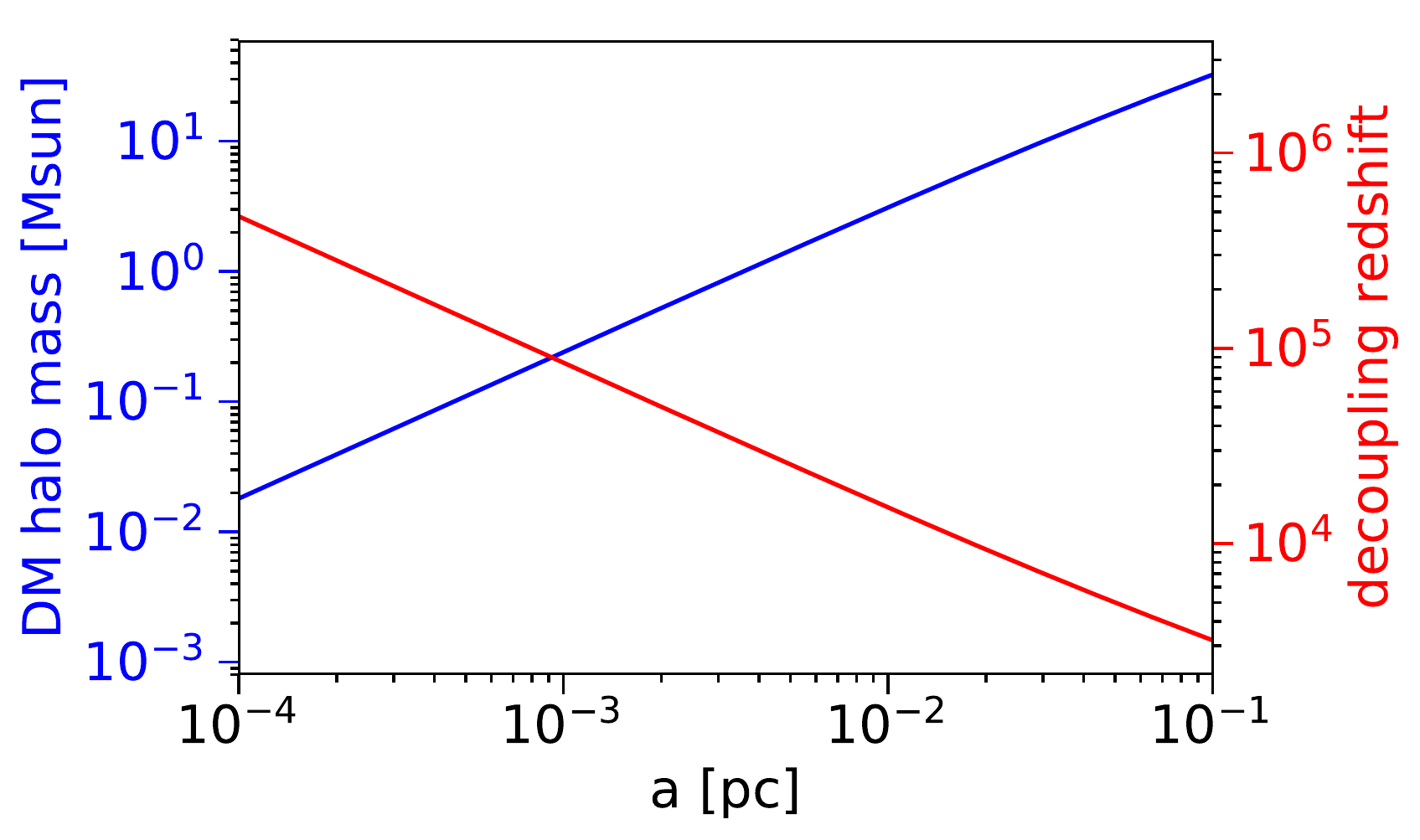}
      \caption{{\bf Decoupling redshift and corresponding mass of the accreted DM halo}. In this plot, we assume that the binary systems are made of $30$ M$_\odot$ PBHs. The quantities are plotted as a function of the semi-major axis, as given by Eq. \ref{eq:z_of_a} and \ref{eq:Mhalo}.} 
      \label{fig:decouplingRedshift}
\end{figure}

It is now important to point out that the time at which a PBH pair decouples from the Hubble flow and forms a binary system depends on the PBH separation. 
The decoupling redshift of a binary with semi-major axis $a$ is given by \cite{Ali-Haimoud:2017rtz}:
\begin{equation}
z_{\rm dec} \, (a) \,=\, 3 f_{\rm PBH} \, z_{\rm eq} \, \frac{\bar{x}^3}{(x(a))^3}
\label{eq:z_of_a}
\end{equation}
where $x(a)$ is defined in Eq.~\eqref{eq:x_of_a}. 

As noted in \cite{Sasaki:2016jop}, the binaries decouple in the radiation era, so we adopt the expression above until the time of matter-radiation equality, i.e. for $z > z_{eq}$. This implies a maximum value of the semi-major axis $a_\mathrm{max}$, set by $z_\mathrm{dec} = z_\mathrm{eq}$ and given by:
\begin{equation}
\label{eq:amax}
\frac{z_{dec}}{z_{eq}} = 3f_{\rm PBH} \left(\frac{\bar{x}}{x(a)} \right)^3 \rightarrow a_{max} = \, \left( \frac{3^{5} \,M_{\rm PBH}}{4 \pi \rho_{eq}}\right)^{1/3}\,.
\end{equation}

We plot in Fig.~\ref{fig:r_tr_and_a} the semi-major axis and the truncation radius as a function of the decoupling redshift, given by Eq.~\eqref{eq:r_tr} and the inverse of Eq.~\eqref{eq:z_of_a}.

We also notice that, from these relations, it is straightforward to realize that wider binaries decouple later (intuitively, for larger separations the gravitational pull takes more time to overcome the Hubble flow that tends to break the pair up), and therefore have the chance to grow a more massive DM halo around each PBH. This behavior is represented in Fig.~\ref{fig:decouplingRedshift}.

\subsection{Self-consistent PBH binary orbits}
\label{sec:selfconsistent}

\begin{figure}[t]
\centering
\includegraphics[width=0.5\textwidth,]{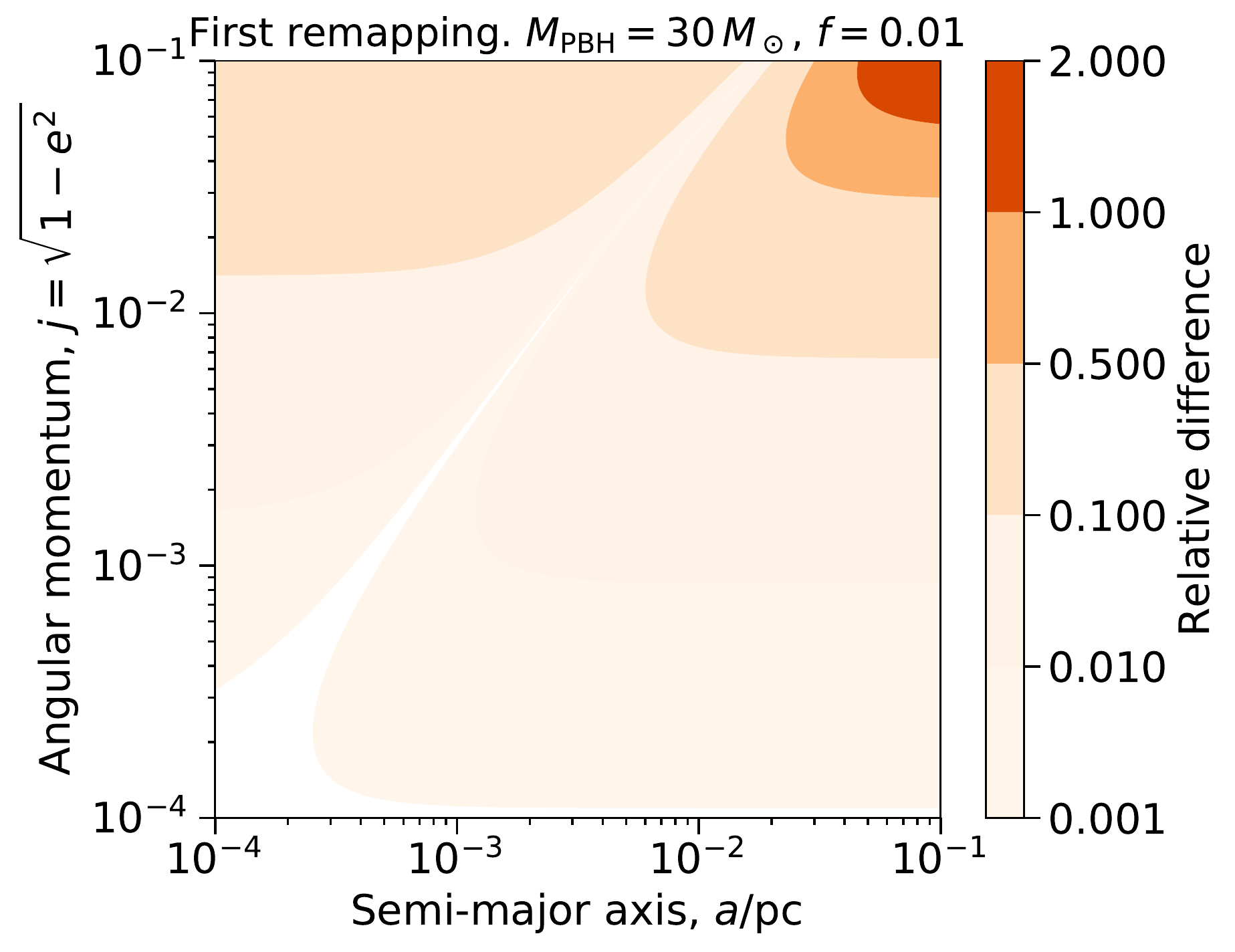}
\caption{{\bf First remapping}. We represent the relative difference between the original PDF in the $(a,j)$ parameter space and the one which takes into account the presence of the mini-halos (as described in Sec. \ref{sec:selfconsistent}). We remark that this initial remapping is mainly based on a rescaling of the mass parameter, and does not take into account the impact of the halos on the BBH orbits, which will be addressed in the next section.}
\label{fig:firstRemapping}
\end{figure}

Before studying the impact of the mini-halos on the merger rate, let us merge the pieces of information presented so far, and estimate how the presence of the mini-halos (discussed in Sec.~\ref{sec:decoupling}) changes the probability distributions in the $(a, j)$ parameter space (described in Sec.~\ref{sec:properties}).

The presence of the halos can be taken into account, at first order, by assuming that the dressed PBHs behave as point masses with $M_{\rm tot} = M_{\rm PBH} + M_{\rm halo}$. 
For a fixed comoving separation, this causes them to decouple {\it earlier} at some redshift $z'$ with a smaller semi-major axis. This implies a rescaling of $x(a)$ in Eq.~\eqref{eq:x_of_a} as follows:
\begin{equation}
 M_{\rm PBH} \, \rightarrow \,  M_{\rm PBH} + M_{\rm halo}(z')\,.
\end{equation}
Moreover, the PDF in Eq.~\eqref{eq:measure} must be rescaled by a factor $(M_\mathrm{tot}/M_\mathrm{PBH})^{3/4}$, arising from the new expression for $\partial x/\partial a$.

  As far as $P(j)$ is concerned, the dominant contribution to the torque comes from large $z$, when the PBHs are  closer together and therefore the forces between them are largest\footnote{This can be seen explicitly in the $s^{-2} \sim z^2$ scaling of the integrand in Eqs.~(13) and (14) of Ref.~\cite{Ali-Haimoud:2017rtz}.}. At large $z$, the PBHs have typically not had time to grow a large DM halo. The late-time growth of the DM halos may increase the torques from other PBHs (and their respective halos), and the DM halo accreted by the nearest neighbors could in principle exchange angular momentum with the binary. However, the mass accretion mostly happens at late times, while the torques are mostly exerted at early times, which means that the angular momentum \textit{per unit mass} (and therefore $j$) are expected to be roughly constant. We therefore assume that the DM halo does not substantially affect the time evolution of the angular momentum and use $P(j)$ given in Eq.~\eqref{eq:Pj} throughout.

In the end, the impact of these corrections is not huge in most of the parameter space: we show in Fig.~\ref{fig:firstRemapping} the relative difference between the original PDF and the ``remapped'' one. Perhaps the largest effect is that it is possible for wider binaries to form when the mass of the halo is included. For binaries decoupling close to $z_\mathrm{eq}$, we can treat the dressed PBH as a point mass $M_\mathrm{tot} \approx M_\mathrm{PBH} + M_\mathrm{halo}(z_\mathrm{eq}) \approx 2 M_\mathrm{PBH}$. This means that the maximum possible semi-major axis $a_\mathrm{max}$, given in Eq.~\eqref{eq:amax}, is increased by a factor of $2^{1/3} \approx 1.26$. This leads to a slight increase in the total number of binaries which can be formed. 

\section{Impact of DM mini-halos on BBH orbits}
\label{sec:simulations}

\subsection{N-body simulations}
\label{sec:Nbody}

In order to assess the impact of DM mini-halos on the orbits of PBH binaries, we use N-body simulations to follow the dynamics of these systems. We use the publicly available \Gadget code \cite{Springel:2005mi} as a gravity-only N-body solver. In this section, we summarise the key features of the simulations, with full details given in Appendix~\ref{sec:Gadget}. The code for setting up and analysing the simulations is publicly available \href{https://github.com/bradkav/BlackHolesDarkDress}{here} \cite{DarkDressCode}. Selected animations are also available \href{https://doi.org/10.6084/m9.figshare.6298397}{here} \cite{Animations}. 

Considering first a single PBH, we set up the surrounding DM halo with a density profile similar to that of Eq.~\eqref{eq:rhoDM} but with a smooth truncation at the truncation radius. The truncation radius and mass of the halo are set based on the semi-major axis of the orbit, as discussed in Sec.~\ref{sec:BBHproperties}. We initialise each DM halo in equilibrium with an isotropic, spherically symmetric velocity distribution obtained using the Eddington inversion formula \cite{2008gady.book.....B}.

We initialise a binary with a given $(a, e)$ as if it consisted of two point masses of $M_\mathrm{tot} = M_\mathrm{PBH} + M_\mathrm{halo}$ each and we begin the simulation during the first in-fall of the PBHs from apoapsis.
We set the softening length for Dark Matter pseudo-particles to be at least a factor of 5 smaller than the distance of closest approach of the PBHs $r_\mathrm{min} = a_i (1-e_i)$. For eccentricities smaller than $e = 0.995$, we use roughly $10^4$ equal-mass DM particles per halo. For eccentricities larger than $e = 0.995$ (requiring a finer resolution), we employ a multi-mass scheme \cite{Zemp:2007nt} using four different masses of DM particles. In this case, we use a total of roughly $4 \times 10^4$ DM particles per halo. We have checked that the density profile of the DM halo is stable down to $r_\mathrm{min}$ on time scales corresponding to the time of the first close passage of the PBH binaries under consideration. Further details are provided in Appendix~\ref{sec:Gadget}. 

We follow the evolution of the binary system until the semi-major axis and eccentricity of the PBH-PBH system have stabilised. 
The final eccentricity and semi-major axis are then estimated from the specific energy $\epsilon$ and specific angular momentum $h$ of the PBH-PBH binary:
\begin{equation}
e = \sqrt{1 + \frac{\epsilon h^2}{2 (G_N M_\mathrm{PBH})^2}}\,,\;\; a = -\frac{G_N M_\mathrm{PBH}}{2\epsilon}\,.
\end{equation}
Here, $\epsilon = \frac{1}{2}v^2 - 2 G_N M_\mathrm{PBH}/r$ and $h = |\mathbf{r} \times \mathbf{v}|$, for PBH separation $\mathbf{r}$ and relative velocity $\mathbf{v}$ \cite{9780471146360}. 

\begin{figure}[t!]
\centering
      \includegraphics[width=\linewidth]{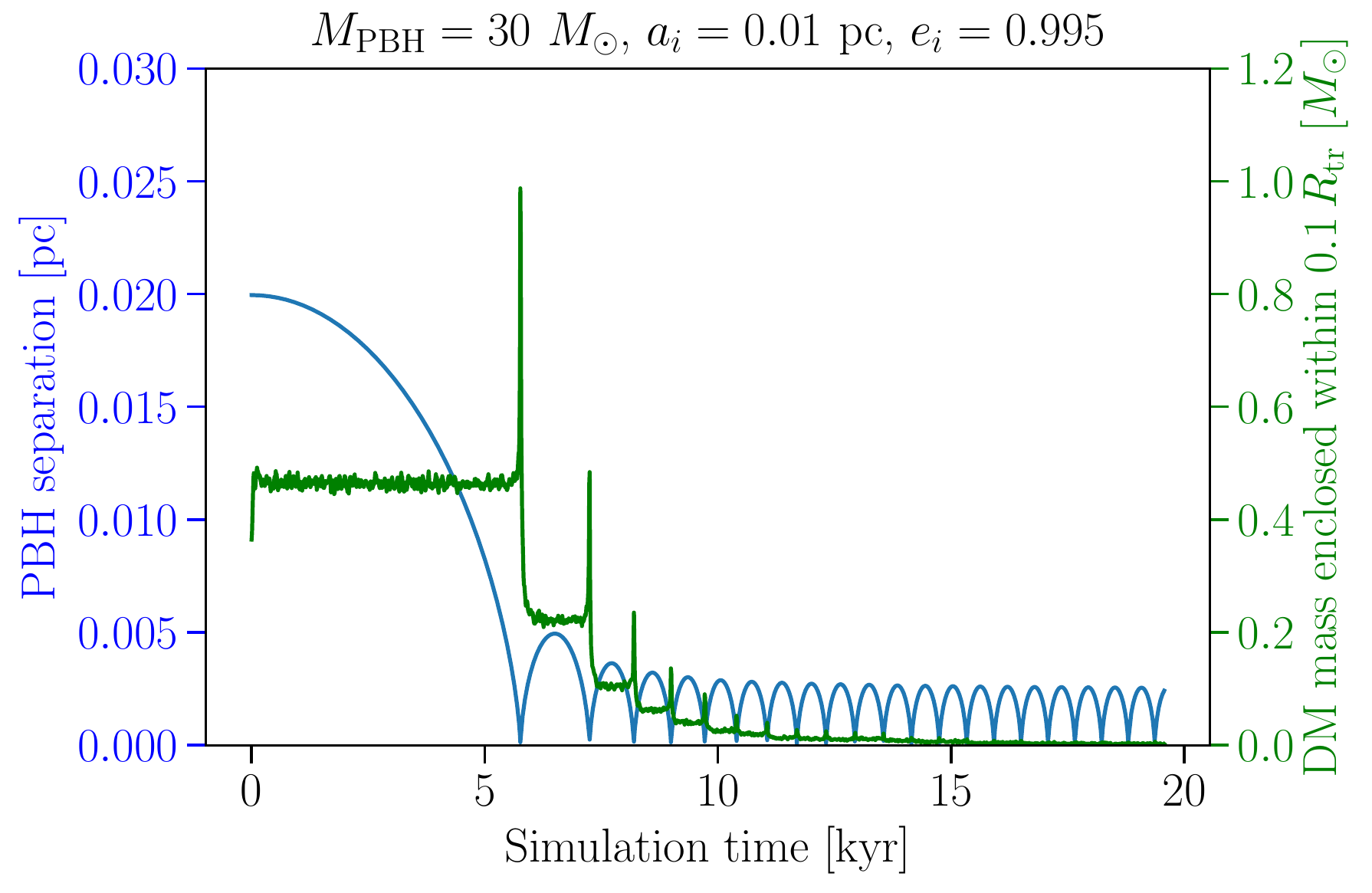}
   \caption{\textbf{PBH separation and retained DM halo mass during a single simulation.} In blue (left axis), we show the separation of the PBHs during a single simulation while in green (right axis) we show the DM mass enclosed within 10\% of the halo truncation radius, $R_\mathrm{tr}$. Here, we simulate $M_\mathrm{PBH} = 30\,M_\odot$ and initial orbital elements $a_i = 0.01\,\mathrm{pc}$ and $e_i = 0.995$. The truncation radius is $R_\mathrm{tr} \approx 4 \times 10^{-3}\,\mathrm{pc}$ and the total DM mass per halo is $3.1\,M_\odot$.}
   \label{fig:PBHseparation}
\end{figure}

\begin{figure}[t!]
\centering
      \includegraphics[width=\linewidth]{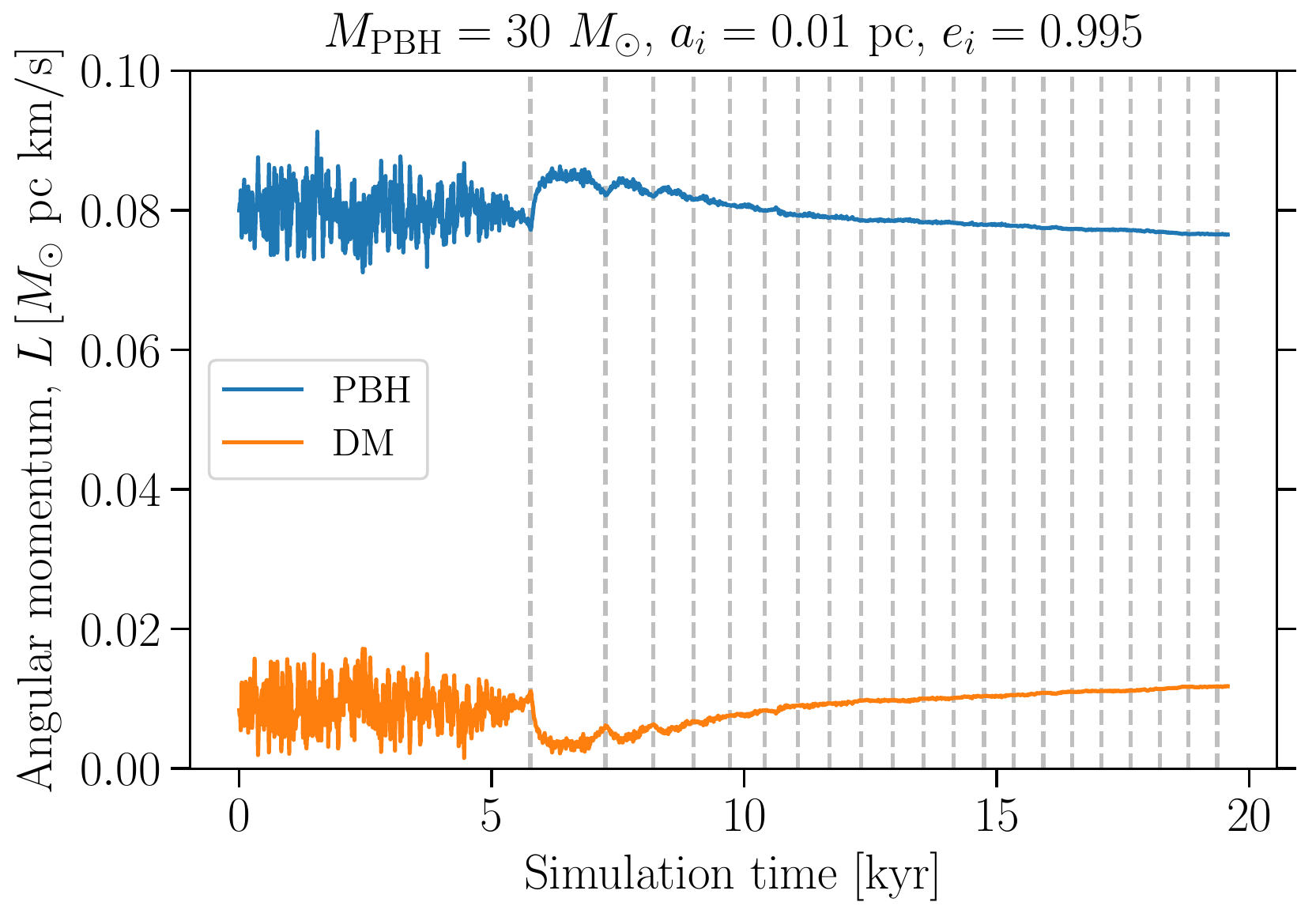}
   \caption{\textbf{Angular momentum of PBHs and DM during a single simulation.} The total angular momentum of the PBH (DM) particles in the simulation is shown in blue (orange). The simulation parameters are as in Fig.~\ref{fig:PBHseparation}. The times at which the PBHs undergo a close passage are marked by grey dashed lines.}
   \label{fig:PBHangmom}
\end{figure}

In Figs.~\ref{fig:PBHseparation}~and~\ref{fig:PBHangmom}, we show the main properties of the binary system during a single simulation, specifically a pair of dressed $30 \,M_\odot$ PBHs with initial orbital elements $a_i = 0.01 \,\mathrm{pc}$ and $e_i = 0.995$. Figure~\ref{fig:PBHseparation} shows the separation of the PBHs as a function of simulation time (blue), as well as the DM mass enclosed with $0.1\,R_\mathrm{tr}$ around one of the PBHs (green). 

During each close passage, the enclosed DM mass jumps by a factor of roughly 2, as the PBH passes through the halo of its companion. 
After the close passage the remaining DM mass is reduced, as a significant fraction of the halo is ejected by the close encounter. This key feature -- feedback between the PBHs and DM halos -- drives the shrinking of the binary orbit. With successive orbits, the DM mass is gradually depleted and the semi-major axis shrinks until it eventually stabilises. This typically take $<\mathcal{O}(10)$ orbits, on timescales $\mathcal{O}(10\,\mathrm{kyr})$.

In Fig.~\ref{fig:PBHangmom}, we plot the angular momentum of the same system. In blue we plot the total angular momentum of the two PBHs, while in orange we plot the total angular momentum of the DM halos. During the first close passage, at $t \sim 5.8 \,\mathrm{kyr}$, there is very little exchange of angular momentum. While dynamical friction acts to slow the PBHs as they pass through the halos, the orbit is almost radial so there is no resulting torque. We note, however, the slowing of the PBHs close to periapsis slightly circularises the orbit.

As the PBHs move away from their first close passage, they then encounter the particles of the disrupted DM halo, which have been ejected with high speed. In this case, dynamical friction acts to accelerate the PBHs and they begin to regain angular momentum. With each successive close passage, however, the effects of dynamical friction with the remaining DM halo particles will slow the PBHs, inducing a torque on the (now more circular) binary. 

For the eccentric $e=0.995$ binary we consider here, the angular momentum of the PBH system at late times is comparable to the initial value. In less eccentric binaries, we have observed that the DM halo can carry away a substantial fraction of the PBH angular momentum. Increasing the eccentricity, on the other hand, typically decreases the amount of angular momentum exchanged between the PBHs and DM halos.

\begin{figure*}[t!]
\centering
	   \includegraphics[width=0.49\linewidth,]{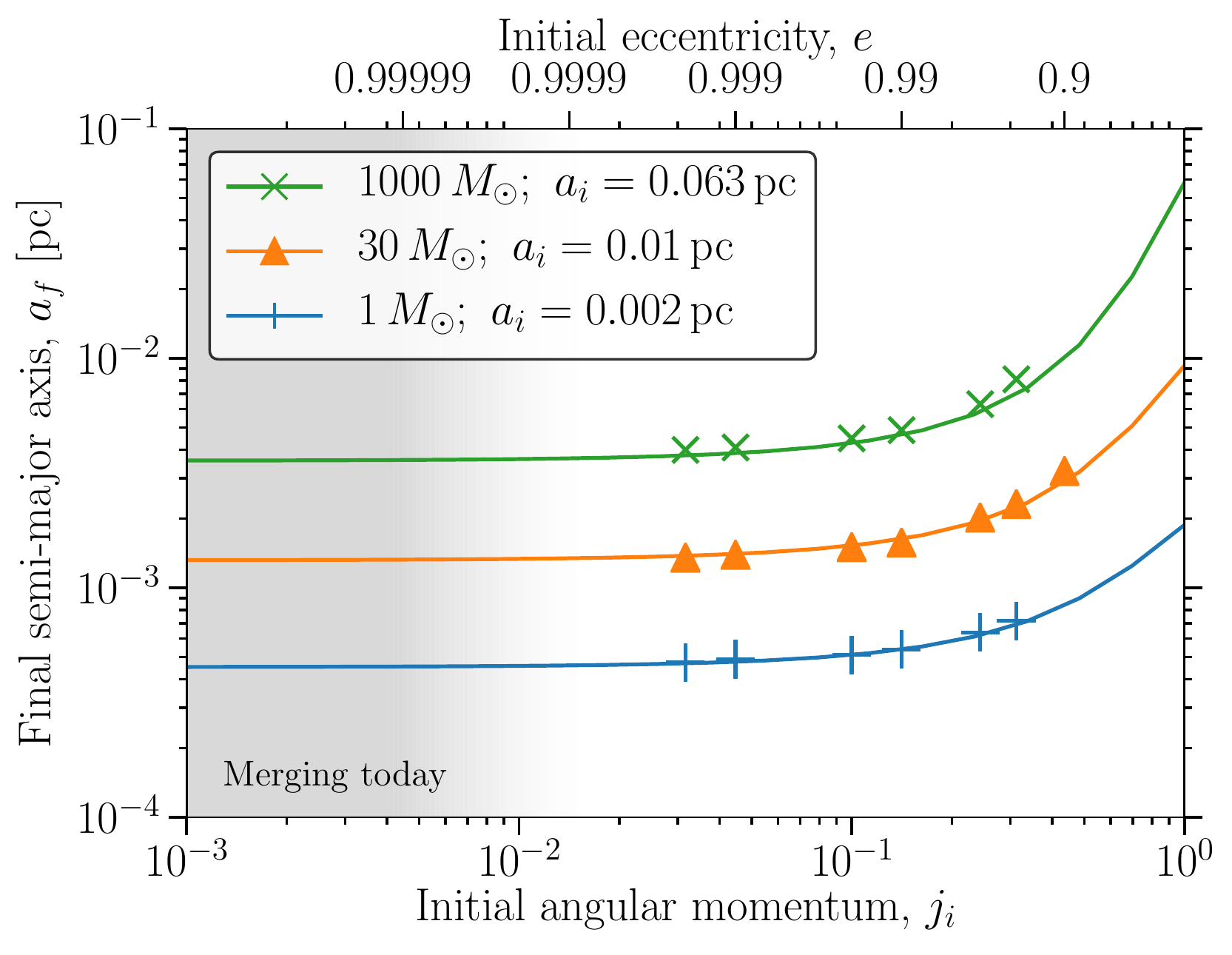}
      \includegraphics[width=0.49\linewidth,]{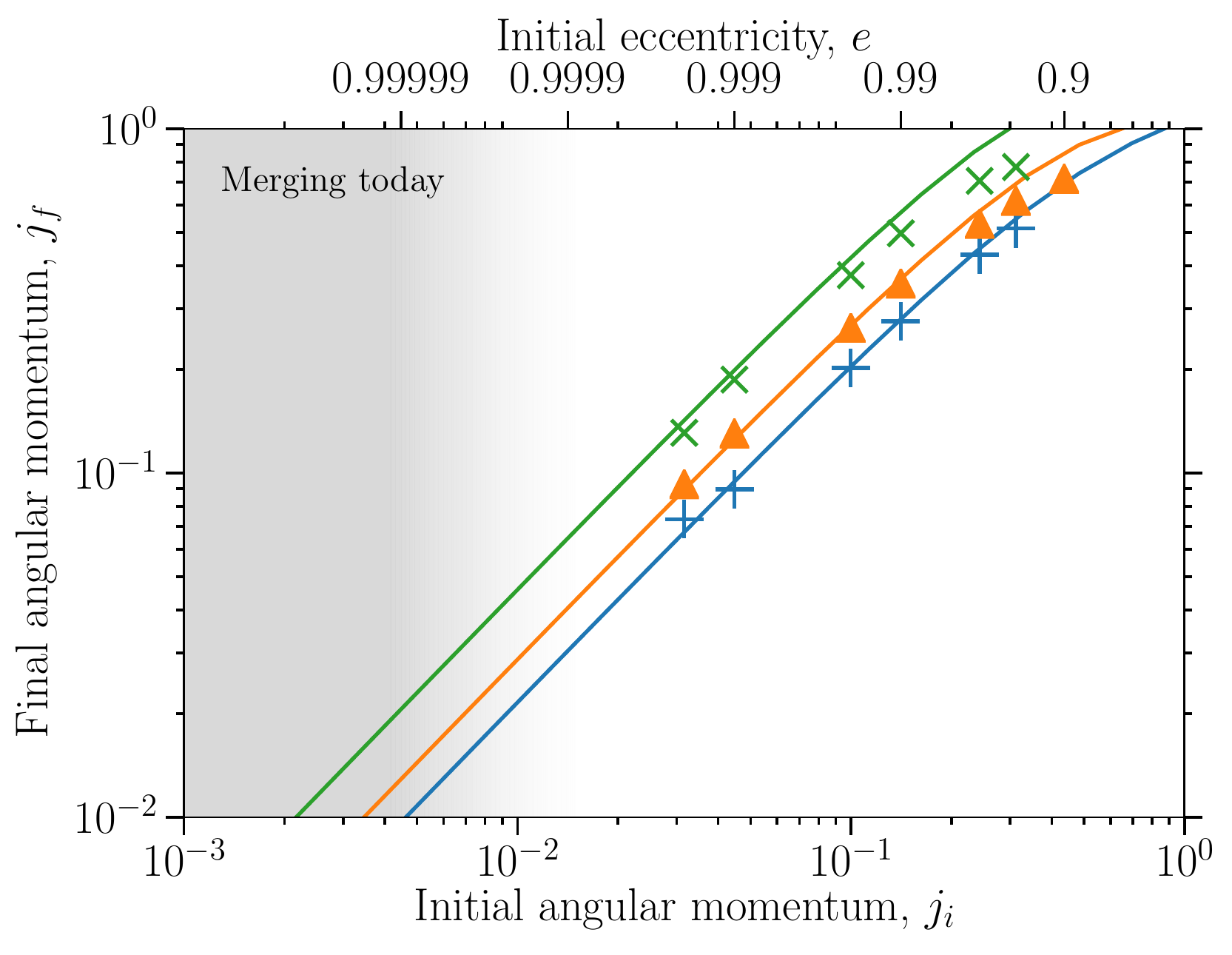}
   \caption{\textbf{Impact of Dark Matter halos on the orbital elements of PBH binaries.} We show the final semi-major axis $a_f$ (\textbf{left}) and final angular momentum $j_f$ (\textbf{right}) of the PBH binaries at the end of our N-body simulations, as a function of the initial angular momentum $j_i$. Each point corresponds to the result of a single simulation run while the solid lines correspond to the analytic estimates which we describe in Sec.~\ref{sec:analytic} (these curves are \textit{not} fit to the data). We show results for three different PBH masses, in each case with a different initial semi-major axis $a_i$. The grey shaded region illustrates typical values of $j$ for which the binaries are expected to merge on timescales of order the age of the Universe. 
   }
   \label{fig:NbodyResults}
\end{figure*}

In Fig.~\ref{fig:NbodyResults}, we show the final semi-major axis $a_f$ and final angular momentum $j_f$ for a number of simulated binary systems. We show results for three PBH masses -- $1\,\,M_\odot$, $30\,\,M_\odot$ and $1000\,\,M_\odot$ -- and in each case we select the most likely initial semi-major axis $a_i$ for binaries merging today (see Fig.~\ref{fig:ProbabilityDistribution}). We see that the final semi-major axis (left panel) is typically smaller than the initial semi-major axis by a factor of $\mathcal{O}(10)$, meaning that the final orbit is much smaller when the DM halo around each PBH is significant. The final orbit is also more circular $j_f > j_i$, as we see in the right panel of Fig.~\ref{fig:NbodyResults}. These two changes - shrinking and circularisation of the binary - have opposing effects on the merger time, Eq.~\eqref{eq:tmerge}, of the binary.

From Fig.~\ref{fig:ProbabilityDistribution}, we see that binaries merging today typically have angular momenta in the range $j = 10^{-3}\text{--}10^{-2}$. We have performed simulations down to $j \approx 0.03$ ($e = 0.9995$) but realistic simulations corresponding to binaries merging today would require around 2 orders of magnitude improvement in spatial resolution in the DM halo (owing to the much smaller close passage distances). As we outline in Appendix~\ref{sec:Gadget}, performing large numbers of such simulations would be computationally infeasible. Instead, in the next section, we use analytic arguments to understand the behaviour of binaries merging today.


\subsection{Analytic results}
\label{sec:analytic}

Guided by the results of our numerical simulations, we now present analytic estimates which capture the key features. As we will see, the resulting expressions are rather simple, but are not trivial to derive without input and validation from N-body simulations (as presented in Sec.~\ref{sec:Nbody}).

\subsubsection{Semi-major axis}
First, we consider the evolution of the semi-major axis of the BBH orbits, incorporating the effects of the DM halos surrounding them using simple energy conservation arguments. Initially, the total orbital energy of the system is given by:
\begin{equation}
E_i^\mathrm{orb} = -\frac{G_N M_\mathrm{tot}^2}{2 a_i}\,,
\end{equation}
where $M_\mathrm{tot} = M_\mathrm{PBH} + M_\mathrm{halo}$ and we have treated each PBH and its halo as a point object. The binding energy of each DM halo, including all DM particles at a distance greater than $r_\mathrm{in}$ from the PBH, is given by:
\begin{equation}
E^\mathrm{bind}(r_\mathrm{in}) = -4 \pi G_N \int_{r_\mathrm{in}}^\infty \frac{M_\mathrm{enc}(r)}{r} \, r^2 \rho_\mathrm{DM}(r) \,\mathrm{d}r \,.
\end{equation}

From the simulations, we see that the work done by dynamical friction unbinds the DM halo, with more of the halo unbound as the distance of closest approach $r_\mathrm{peri} = a_i (1-e_i)$ decreases. We assume that each PBH maintains a halo of radius $r_\mathrm{min}/2$, with DM particles further away than this being completely unbound. The final orbital energy of the binary is then given by:
\begin{equation}
E_f^\mathrm{orb} = -\frac{G_N M_f^2}{2 a_f}\,,
\end{equation}
where $M_f = M_\mathrm{PBH} + M_\mathrm{halo}(r < r_\mathrm{min}/2)$. 
The final semi-major axis $a_f$ is then obtained (for a given $r_\mathrm{min}$ and therefore a given $j_i = \sqrt{1-e_i^2}$) from energy conservation,
\begin{equation}
E_i^\mathrm{orb} + 2 \,E^\mathrm{bind}(r_\mathrm{min}/2) = E_f^\mathrm{orb}\,.
\label{eq:energy_conservation}
\end{equation}
The final semi-major axis calculated in this way can be written explicitly as follows:
\begin{equation}
a_f (a_i) \,=\, \, \frac{G_N M_{f}^2 a_i}{G_N M_{tot}^2 + 4 a_i E^\mathrm{bind}(r_\mathrm{in})}\,.
\label{eq:afinal}
\end{equation}
We show this result in the left panel of Fig.~\ref{fig:NbodyResults} as solid lines for the three different scenarios. For circular orbits ($j_i \rightarrow 1$) there is little change in the semi-major axis as the PBHs do not pass within each other's DM halos\footnote{Note that over longer periods, tidal effects would be expected to disrupt the two halos. We are interested in much more eccentric binaries and so we do not consider this effect further.}. For increasingly eccentric binaries, more and more of the DM halo is stripped, reducing the final orbital energy of the PBH pair and therefore the final semi-major axis. At high eccentricity ($j_i \ll 1$), almost all of mass of each DM halo is stripped; almost all of the halo binding energy is converted to orbital energy and decreasing $j_i$ further has no impact on the final semi-major axis.

\begin{figure}[t!]
\centering
   \includegraphics[width=0.95\linewidth,]{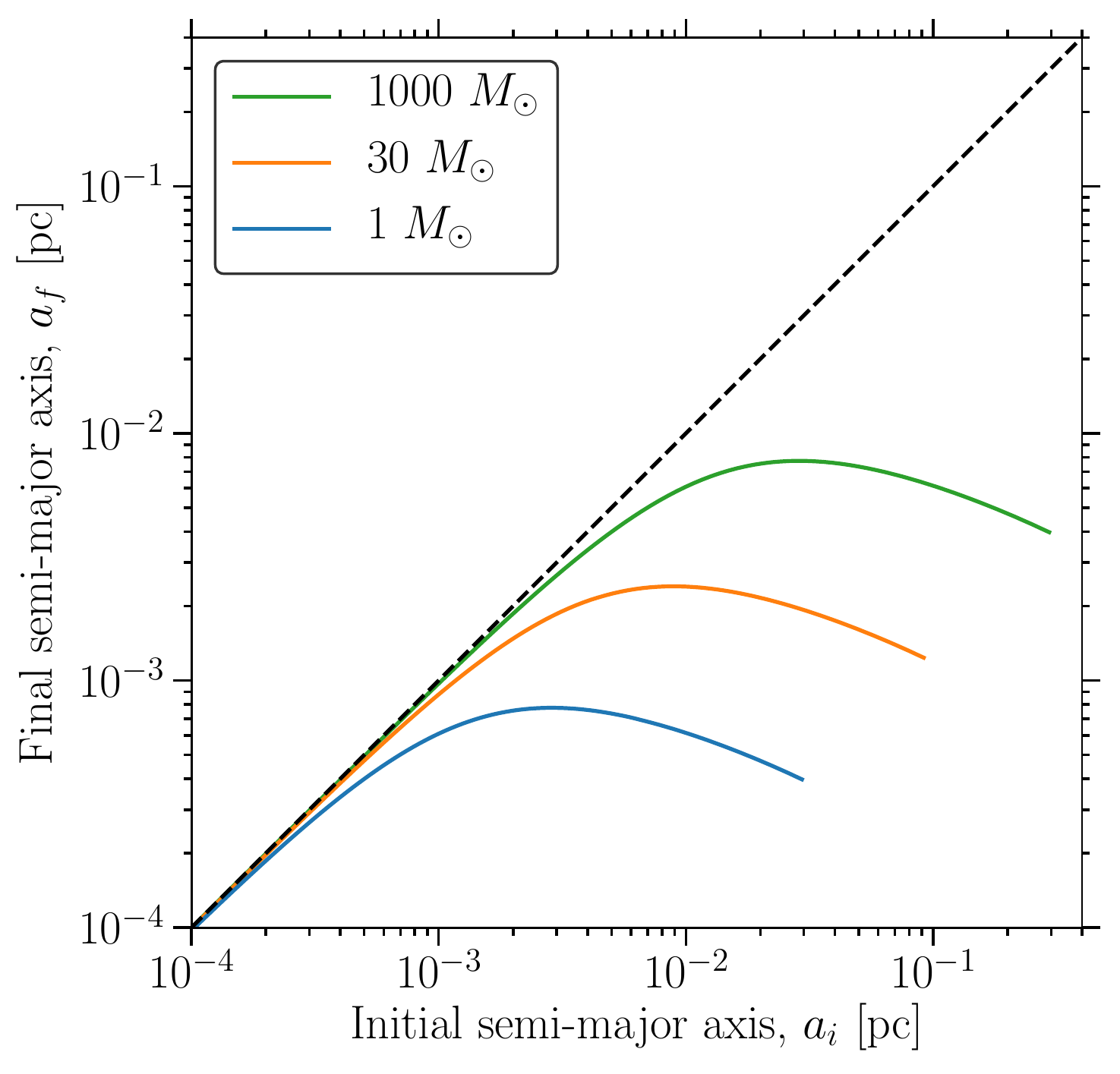}
   \caption{\textbf{Impact of DM halos on the semi-major axis of highly eccentric PBH binaries.} Final semi-major axis of PBH binaries after their local DM halos have been disrupted and unbound, following the analytic prescription of Sec.~\ref{sec:analytic}. We show results for 3 different PBH masses and assume the DM density profile given in Eq.~\eqref{eq:rhoDM}. The black dashed line corresponds to $a_f = a_i$.}
   \label{fig:semimajoraxis}
\end{figure}

In Fig.~\ref{fig:semimajoraxis}, we show the analytic estimate of $a_f$ as a function of $a_i$ for binaries with PBH masses of $1 \,\,M_\odot$, $30 \,\,M_\odot$ and $1000 \,\,M_\odot$. In this case, we assume a DM density profile given by Eq.~\eqref{eq:rhoDM} and assume that the entire DM halo of each PBH is stripped, which is valid for highly eccentric orbits. For small orbits ($a_i \lesssim 10^{-4} -10^{-3} \,\,\mathrm{pc}$) we find little change in the semi-major axis. This is because these binaries decouple from the Hubble flow early and have not had time to grow a substantial DM halo. The impact of the DM halo increases with increasing semi-major axis, as the binary decouples later and the size of the halo at decoupling grows. 

\subsubsection{Angular Momentum}

As in the case of the semi-major axis, we can use conservation arguments to estimate the final dimensionless angular momentum $j$ of the orbits after the effects of the DM halo have been taken into account.

The dimensionful angular momentum $L$ for a binary of two point masses $M$ is given by:
\begin{equation}
\label{eq:AngMom}
L^2 = \frac{1}{2}G_N M^3 \, a \, j^2\,.
\end{equation}
As we have seen from the N-body simulations in the previous section (in particular Fig.~\ref{fig:PBHangmom}), for very eccentric orbits there is very little exchange of angular momentum between the PBHs and the DM particles. This can be understood from the fact that for large eccentricity the orbits are almost radial. This means that there is very little torque acting on the PBHs, despite the large dynamical friction force. At the distance of closest approach, the PBH velocity is perpendicular to PBH separation and the DM density is highest, in which case we might expect a large torque. However, this is also the point in the orbit where the PBHs have the highest velocity, suppressing the dynamical friction force \cite{Chandrasekhar1943a}. As we see from our N-body results, the latter effect dominates and very little angular momentum is exchanged.

As discussed in Sec.~\ref{sec:BBHproperties}, we are interested in highly eccentric binaries $j \lesssim 10^{-2}$ (corresponding to $e \gtrsim 0.9999$) which are expected to merge today. In this case then, we may assume that there is no angular momentum exchange, in which case the angular momentum of both the PBHs and the DM halos are \textit{separately} conserved. From this, it holds that
\begin{equation}
L^2 = \frac{1}{2}G_N M_\mathrm{PBH}^3\, a\, j^2\,,
\end{equation}
is conserved and therefore that:
\begin{equation}
\label{eq:jf}
j_f = \sqrt{\frac{a_i}{a_f}}j_i \qquad \text{for } j \ll 1\,.
\end{equation}
Combined with the prescription for calculating the final semi-major axis, this allows us to calculate the final angular momentum of the PBH binaries.

In the right panel of Fig.~\ref{fig:NbodyResults}, we plot as solid lines the estimates of $j_f$ (given by Eq.~\eqref{eq:jf}), which agree well with the N-body simulation results at small $j_i$. For large $j$, the final angular momentum is smaller than this estimate would suggest. In this case, the more circular orbits lead to angular momentum exchange between the PBHs and their DM halos; the torque from dynamical friction reduces the angular momentum of the PBH binary. The conservation of angular momentum of the PBH binary is not an intrinsic property of the system then, but only a special quality of the most eccentric orbits, relevant for mergers today.

\subsubsection{Merger times}
\label{merger_times}

With the results of the previous sections at hand, we can now calculate the final merger time for a binary (Eq.~\eqref{eq:tmerge}), given its initial orbital elements. 

We note here that the merger time scales $t_\mathrm{merge} \propto a^4 j^7$, while the conserved angular momentum of the PBH binary scales as $L^2 \propto a j^2$: This indicates that, despite the strong scaling of the merger time with $a$ and $j$, the final merger time will not be changed substantially by the DM halo. Indeed, substituting Eq.~\eqref{eq:jf} into Eq.~\eqref{eq:tmerge}, we find that,
\begin{equation}
t_f = \sqrt{\frac{a_i}{a_f}}\,t_i\,,
\label{eq:merger_time_final}
\end{equation}
where $t_i$ and $t_f$ are the initial and final merger times of the binary, before and after the impact of the DM halo are taken into account. As we see in Fig.~\ref{fig:semimajoraxis}, the semi-major axis is typically not reduced by more than a factor of 10, meaning that the merger time is unlikely to be reduced by more than a factor of a few.

\section{Merger Rates and Constraints on the PBH density}
\label{sec:results}

We can now combine the various findings described in the previous sections in order to compute the impact of DM mini-halos on the primordial BBH merger rate and the corresponding LIGO limit on the PBH fraction.

Let us recap in detail the prescription we follow:
\begin{itemize}

\item We begin with the distribution of orbital elements $(a, e)$, or equivalently $(a, j)$, for PBH binaries in the early Universe, as described in Sec.~\ref{sec:selfconsistent}.

\item For a PBH binary with a given semi-major axis, we estimate the redshift $z_\mathrm{dec}$ at which the pair decouples from the Hubble flow, and calculate the DM halo mass accreted at that redshift. 

\item We compute the final semi-major axis and eccentricity of the binary adopting the relations derived above -- summarized by Eqs.~\eqref{eq:afinal}~and~\eqref{eq:jf} -- in order to calculate the new distribution of orbital elements $(a, e)$. 

\item Once this remapping is performed, we calculate the corresponding distribution of merger times and, eventually, we obtain: {\bf 1)} The merger rate {\it today} of PBH binaries formed in the early Universe (to be compared to the one derived by assuming the original distribution of orbital elements derived in \cite{Ali-Haimoud:2017rtz} and given by Eq.~\eqref{eq:Paj}); {\bf 2)} The corresponding limit on the fraction of DM in PBHs.
\end{itemize}

Let us now present and discuss the details of this procedure, and the two main results of the calculation.

\subsection{Merger Rate Today}

\begin{figure}[t]
\centering
   \includegraphics[width=\linewidth,]{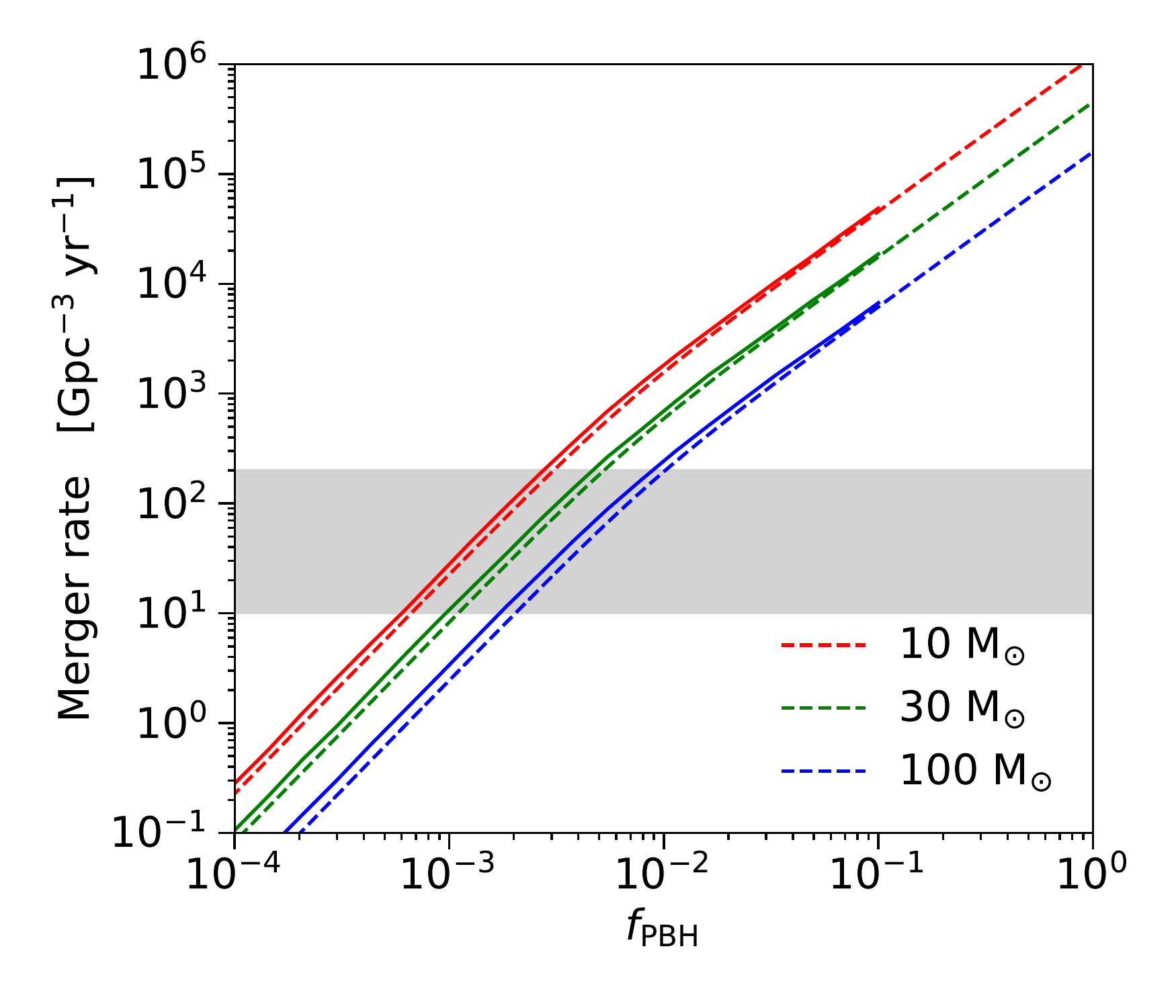}
   \caption{
   \textbf{Primordial Black Hole merger rate,  averaged between $z = 0$ and $z = 1$, as a function of the DM fraction.}   
   {\it Dotted lines:} Merger rate for the ``naked'' PBH binary distribution derived in \cite{Ali-Haimoud:2017rtz}. 
   {\it Solid lines}: Merger rate for the ``dressed'' PBH binary distribution, with the effect of dynamical friction taken into account, as derived in the present work.
   {\it Gray band}: Merger rate inferred by the LIGO and Virgo collaboration, from \cite{Abbott:2017vtc}.
   }
   \label{fig:ratePlot}
\end{figure}

The merger rate of primordial BBHs at present time\footnote{Note that $\mathcal{R}$ is the comoving merger rate density {\it in the source frame.}} is given by:
\begin{equation}
\mathcal{R}_0 = n_\mathrm{PBH} P(t_\mathrm{merge} = t_\mathrm{univ})\,,
\end{equation}
where $n_\mathrm{PBH}$ is the comoving number density of PBHs and $t_\mathrm{univ} \approx 13.7 \,\,\mathrm{Gyr}$ is the age of the Universe. 
However, since LIGO probes mergers approximately in the range $z \in [0, 1]$, we consider the rate averaged over redshift:
\begin{equation}
\langle \mathcal{R} \rangle = n_\mathrm{PBH} \int_{0}^{1} P(t[z])\,\mathrm{d}z \,.
\end{equation}
We now compute the probability distribution of the merger time for both the original PDF given by Eq.~\eqref{eq:Paj}, and for the remapped one, that takes into account the impact of DM dresses.

In the former case, the computation can be carried out analytically by performing a change of variables and a marginalization over the semi-major axis as follows:
\begin{equation}
\label{eq:Pt}
P(t) \, = \, \int_{a_\mathrm{min}}^{a_\mathrm{max}}{ P(a, j(a,t)) \,\left(\frac{{\rm d} j}{{\rm d} t}\right) \,{\rm d}a }\,,
\end{equation}
where $j(a,t)$ is obtained by inverting Eq.~\eqref{eq:tmerge}.
%

In the latter case, we perform a numerical estimate as follows. 
We first sample the original PDF by means of an affine-invariant MCMC ensemble sampler \cite{ForemanMackey:2012ig} and obtain a collection of $\sim 10^5$ points in the $(a, j)$ parameter space. We then apply the remapping prescriptions presented in the previous section (Eq. \ref{eq:afinal} and Eq. \ref{eq:jf}) to this set of points, and eventually determine the final distribution of merger times associated to the remapped points.

We show the result in Fig.~\ref{fig:LIGO_limit}. As argued in Sec.~\ref{merger_times}, the merger time distribution is not strongly affected by the remapping, despite the significant changes in the properties of the binaries, and the strong scaling of the coalescence time with $a$ and $j$. This highly non-trivial result mainly stems from the fact that the shrinking and the circularisation of the binaries (that affect the merger time in opposite directions) are not independent, given the separate conservation of both the PBH and DM dress angular momentum, derived in the previous sections.

\subsection{LIGO/Virgo upper limit}

We now turn to the upper limit on the PBH fraction, which can be obtained by comparing the merger rate predicted for a given $M_\mathrm{PBH}$ and $f_\mathrm{PBH}$ with the upper limit determined by the LIGO experiment. These upper limits are obtained assuming that the merger rate is constant as a function of comoving volume and time in the source frame \cite{Abbott:2016drs,Abbott:2017iws}. For the PBH binaries we consider, however, the merger rate is not constant in time in the source frame, as the distribution of merger times given by Eq.~\eqref{eq:Pt} is not flat.

The \textit{effective} merger rate\footnote{This is the merger rate constant in time in the source frame which would produce the same number of above-threshold events in LIGO as the true time-dependent merger rate $\mathcal{R}(z)$.} which would be measured by LIGO is therefore given by:
\begin{equation}
\label{eq:RLIGO}
\mathcal{R}_\mathrm{LIGO} = n_\mathrm{PBH} \frac{\int S(z) P(t[z])\,\mathrm{d}z}{\int S(z) \,\mathrm{d}z}\,,
\end{equation}
where $S(z) = \mathrm{d}\langle VT \rangle/\mathrm{d}z$ is the space-time sensitivity of LIGO as a function of redshift and depends on the mass of the merging BHs. For $M_\mathrm{BH} = 10,\,20,\,40\,\,M_\odot$, we obtain $S(z)$ from Fig.~7 of Ref.~\cite{Abbott:2016drs}. For $M_\mathrm{BH} = 100,\,200,\,300\,\,M_\odot$, we assume that the overall shape of $S(z)$ does not change substantially from the $40\,\,M_\odot$ case. For a given BH mass, we then rescale $S(z)$ such that the maximum redshift to which LIGO is sensitive corresponds to the horizon distance (Fig.~1 of Ref.~\cite{Abbott:2017iws}) for that mass. We then adjust the normalisation to give the correct value of the space-time volume sensitivity $\langle VT \rangle = 2.302/\mathcal{R}_{90\%}$ (Tab.~1 of Ref.~\cite{Abbott:2017iws}).\footnote{As described in Ref.~\cite{Abbott:2017iws}, the LIGO analysis to search for intermediate mass BHs ($M_\mathrm{BH} \gtrsim 100 \,M_\odot$) is different to that for lighter BH mergers. This is because for intermediate mass BH mergers, only part of the merger and ring-down appear in the LIGO frequency band. This means that rescaling $S(z)$ from $40 \,M_\odot$ up to $300 \,M_\odot$ is not strictly correct. However, the method we use should capture the broad redshift dependence of the sensitivity.}

As in the previous section, we calculate $\mathcal{R}_\mathrm{LIGO}$ by Monte Carlo sampling, in this case weighting each sample by the sensitivity $S(z)$. The LIGO upper limit on the PBH fraction is then obtained by finding the value of $f_\mathrm{PBH}$ for which $\mathcal{R}_\mathrm{LIGO} = \mathcal{R}_{90\%}$.

\begin{figure}[t]
\centering
   \includegraphics[width=\linewidth,]{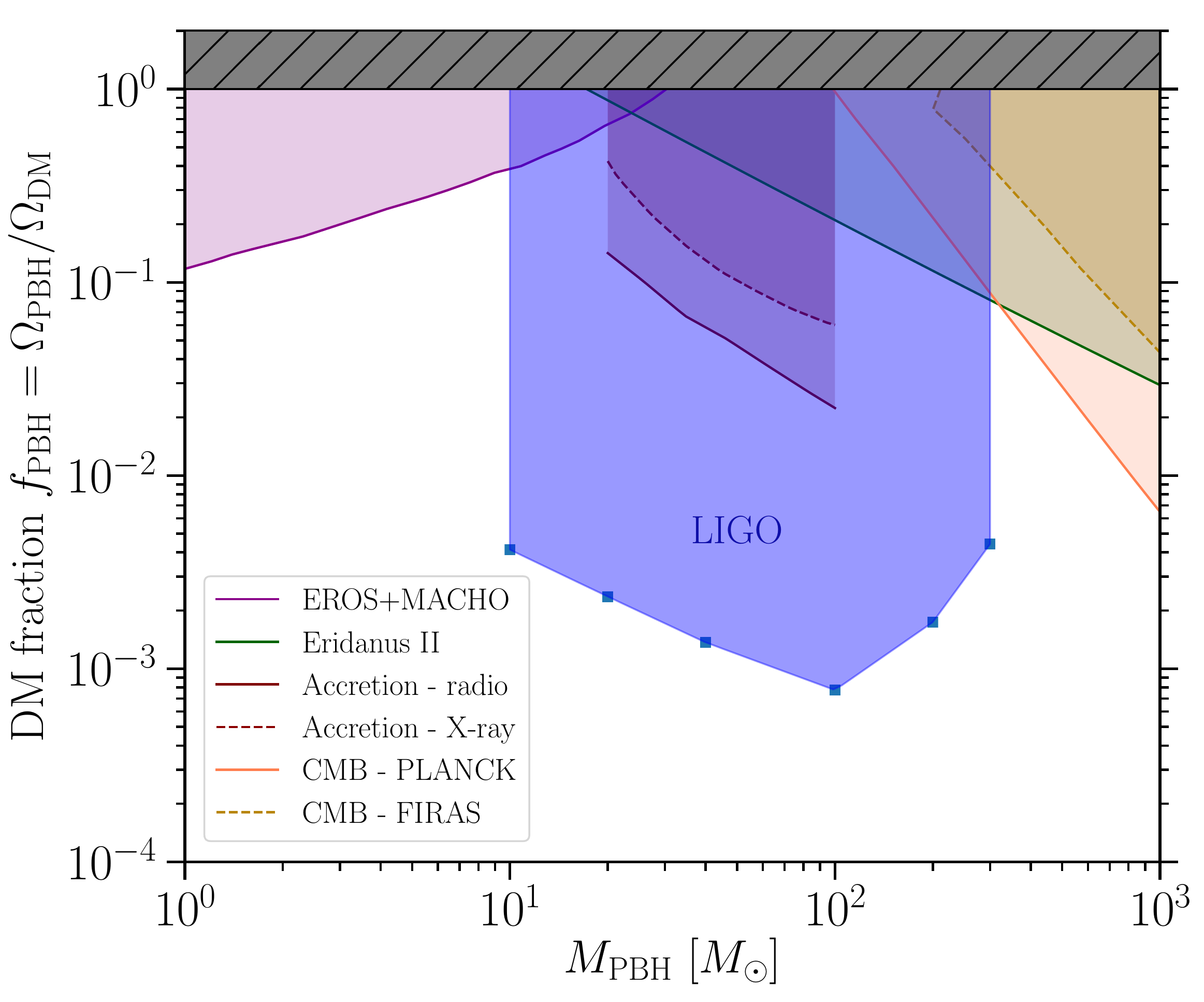}
   \caption{\textbf{Constraints on the fraction $f_\mathrm{PBH}$ of Dark Matter in primordial black holes (PBHs).} Constraints from the LIGO/Virgo merger rate including the effects of local DM halos on PBH binaries (this work) are shown as the dark blue shaded region and labelled `LIGO'. Complementary constraints on PBHs are also shown, with details given in the text. We assume a monochromatic PBH mass function throughout.}
   \label{fig:LIGO_limit}
\end{figure}

In Fig.~\ref{fig:LIGO_limit}, we plot the limits on $f_\mathrm{PBH}$ for the PBH masses listed above, including the effects of the DM dress on the binary evolution. We also show a number of additional constraints on PBHs (assuming mono-chromatic mass functions) in the range $M_\mathrm{PBH} \in [1, 1000]\,M_\odot$. Micro-lensing observations from the MACHO \cite{Allsman:2000kg} and EROS \cite{Tisserand:2006zx} collaborations place constraints on PBH masses up to $30 \,M_\odot$. The presence of PBHs may also disrupt wide binaries \cite{Monroy-Rodriguez:2014ula} and stellar clusters in dwarf galaxies \cite{Brandt:2016aco,Koushiappas:2017chw}. In Fig.~\ref{fig:LIGO_limit}, we show the limit coming from observations of stellar clusters in Eridanus II \cite{Brandt:2016aco}. The accretion of baryons onto PBHs in the inner Milky Way would cause them to radiate: a comparison with known radio and X-ray sources yields constraints at the 1\%--10\% level for PBHs between 20 and 100 $M_\odot$ \cite{Gaggero:2016dpq}. Finally, accretion onto PBHs may distort the CMB spectrum and affect CMB anisotropies \cite{Carr1981}: we show resulting constraints from COBE/FIRAS \cite{Blum:2016cjs,Clesse:2016ajp} and PLANCK \cite{Ali-Haimoud:2016mbv}, using conservative assumptions on accretion onto PBHs in the early Universe. We note that, in general, all constraints on the PBH fraction are subject to a range of uncertainties and caveats (see e.g.~Refs.~\cite{Hawkins:2015uja,Li:2016utv,Garcia-Bellido:2017xvr,Poulin:2017bwe}).

The limits we derive here constrain the PBH fraction to be no more than $4 \times 10^{-3}$ in the mass range $M_\mathrm{PBH} \in [10, 300]\,M_\odot$. The strongest constraints are for $M_\mathrm{PBH} = 100\,M_\odot$ (where the LIGO sensitivity peaks),  giving $f_\mathrm{PBH} < 8 \times 10^{-4}$. These limits on the PBH fraction from the observed merger rate are the most stringent in this mass range, improving on constraints from Galactic radio and X-ray emission \cite{Gaggero:2016dpq} by over an order of magnitude.

\section{Discussion}
\label{sec:discussion}
The presence of the dark matter dress makes the limits on $f_\mathrm{PBH}$ (dark blue shaded region labelled `LIGO' in Fig.~\ref{fig:LIGO_limit}) a factor of 2 stronger than those derived in the case of naked black holes~\cite{Ali-Haimoud:2017rtz}. Roughly 10-20\% of this improvement comes from the inclusion of the DM halos in the decoupling calculations, described in Sec.~\ref{sec:selfconsistent}, which slightly distorts the distribution of orbital elements of PBH binaries. A further $\sim$20\% comes from the impact of the DM halo on the orbit of the PBH binaries, as discussed in Sec.~\ref{sec:simulations}. Finally, around a 50\% increase in the merger rate comes from including the full redshift-dependence of the LIGO sensitivity; there are more binaries with a shorter merger time (corresponding to a higher redshift), increasing the effective merger rate which would be measured by LIGO. 

These results show that local DM halos have a relatively small effect on the merger rates of PBH binaries, despite drastically changing the size and shape of their orbits. For the widest orbits the merger time may be reduced by an order of magnitude, but integrated over the entire population of binaries, the increase in the merger rate today is an $\mathcal{O}(10\%)$ effect. This effect is comparable in size to uncertainties in the detectors' amplitude calibration, which introduces an uncertainty of about 18\% in the upper limit on the merger rate \cite{Abbott:2017iws}. This work therefore increases the robustness of LIGO limits on the PBH fraction (such as those of Refs.~\cite{Sasaki:2016jop,Ali-Haimoud:2017rtz}, as well as those presented here).

A key aspect of our results is that binary systems formed in the early Universe survive the growth of local DM halos. However, there are several ways in which the properties of PBH binaries may also be altered~\cite{Ali-Haimoud:2017rtz}: 
\begin{itemize}
\item They may interact with a circumbinary accretion disk of baryons \cite{Hayasaki:2008eu,Hayasaki:2009ug,Macfadyen:2006jx,Tang:2017eiz}. As we have seen in our simulations, catastrophic feedback between the PBHs and their DM halos is a key feature of the binaries. Therefore, it seems likely then that dedicated numerical simulations will be required to understand how baryonic accretion affects PBH binaries.
\item PBH binaries may be disrupted by interactions with the smooth DM halo or with other PBHs. Binaries formed from dressed PBHs are however typically smaller in size after having unbound their DM halos. From Fig.~\ref{fig:semimajoraxis}, we see that the maximum semi-major axis for binaries of $30\,M_\odot$ PBHs is reduced from $\sim 10^{-1}\,\mathrm{pc}$ to $\sim 2 \times 10^{-3}\,\mathrm{pc}$ once the effects of local DM halos are taken into account. Since smaller binaries are less likely to be disrupted by interactions with the smooth DM halo or with other PBHs, and since naked PBH binaries are not expected to be significantly disrupted~\cite{Ali-Haimoud:2017rtz}, we conclude that dressed PBH binaries should not be affected by mergers into larger virialized DM halos
\item The binary may experience dynamical friction from DM enclosed within the orbit or accreted after decoupling. In this case, the accretion cannot be modeled as in Sec.~\ref{sec:decoupling} as the time-dependent, non-spherical nature of the potential must be accounted for. This is likely to require dedicated simulations and we defer this to future work. However, we expect that DM infalling onto the binary will only be loosely bound and will therefore not dramatically affect the final semi-major axis  and merger time of the binary.

\end{itemize}

We have focused on PBHs characterized by a monochromatic mass function and uniform distribution in space, while realistic scenarios of PBH formation would naturally lead to a PBH population with a range of masses \cite{Kuhnel:2015vtw}, and a significant clustering may be present in the spatial distribution \cite{Chisholm:2005vm}. 
Concerning the former point, the formation of PBH binaries from a population with an extended mass function has recently been studied in Refs.~\cite{Chen:2018czv,Raidal:2017mfl}. In general, constraints may be weakened or strengthened relative to the monochromatic case \cite{Green:2016xgy,Bellomo:2017zsr,Kuhnel:2017pwq,Carr:2017jsz,Lehmann:2018ejc} (depending on the shape of the mass function), so a dedicated reanalysis would be required in this case. 
The analytic treatment we have developed in Sec.~\ref{sec:analytic} could be straightforwardly extended to PBHs of different masses, and we leave this for future studies.
As for the latter point, we expect that significant clustering would further increase the merger rate, making the constraints even stronger (see however Ref.~\cite{Ali-Haimoud:2018dau}). Also in this case, we leave a detailed and quantitative study for future work.

\section{Conclusions}
\label{sec:conclusions}
In this work, we have explored the impact of Dark Matter (DM) halos around black holes (BHs) on the properties of binaries and their merger rates. We have focused on primordial black holes (PBHs), forming binaries in the early Universe, though our techniques are more generally applicable to other merging systems. In the case of PBHs, the growth of DM halos in the radiation-dominated era is a generic prediction and so their impact must be properly included.

We have performed N-body simulations of orbiting PBHs and their respective DM halos, finding that close passages between the BHs tend to disrupt and unbind the DM particles, exchanging orbital energy for gravitational binding energy of the halos. For the most eccentric binaries, relevant for merger events today, we find that little angular momentum is exchanged between the BHs and the DM halos. The results of these simulations have allowed us to determine simple, analytic expressions for how the semi-major axis and eccentricity of binaries change after the disruption of the DM halos. 

Using these relations, we have calculated the distribution of merger times for PBH binaries and the corresponding merger rate which would be observed at LIGO, as a function of the PBH mass $M_\mathrm{PBH}$ and fraction $f_\mathrm{PBH}$. In order not to exceed the limits on BH merger rates set by the LIGO and Virgo collaborations, we set the most stringent limits on the PBH fraction in the mass range $10\text{--}300\,M_\odot$. For PBHs of mass $100 \, M_\odot$, for example, we require $f_\mathrm{PBH} < 8 \times 10^{-4}$. 

These constraints are stronger than the potential limits suggested in Ref.~\cite{Ali-Haimoud:2017rtz} (also based on LIGO merger rates), but still within a factor of $2$. This indicates that, while DM halos around PBHs can substantially alter the size and shape of PBH binary orbits, they lead to only an $\mathcal{O}(10\%)$ effect on the merger rate of PBH binaries today. This result strengthens the case that PBH binaries should survive until today, placing LIGO/Virgo bounds on $f_\mathrm{PBH}$ on more solid ground.

The techniques we have employed are also more generally applicable to astrophysical black holes with a dark matter dress.
A particularly interesting application concerns the analysis of the gravitational waves signal emitted by dressed astrophysical binary black holes. It has been suggested that the presence of dark matter would modify the dynamics of the merger, and induce a potentially detectable dephasing in the waveform \cite{Eda:2013gg,Eda:2014kra}. This conclusion is however based on the assumption that there is no dynamical effect on the dark matter, whose distribution is kept constant with time. We leave the analysis of these systems, and of the ensuing gravitational wave emission, to a future publication. 

\begin{acknowledgements}
We thank Joe Silk for stimulating discussions on PBH mergers, as well as Sarah Caudill and Christopher Berry for helpful discussions on LIGO sensitivities. We also thank Alfredo Urbano, Marco Taoso and Joe Silk for comments on a draft of this manuscript.
Where necessary, we have used the publicly available WebPlotDigitizer \cite{WebPlotDigitizer} to digitise plots.
We thank SURFsara (\href{www.surfsara.nl}{www.surfsara.nl}) for the support in using the Lisa Compute Cluster.
BJK acknowledges funding from the Netherlands
Organization for Scientific Research (NWO) through the VIDI research program ``Probing the Genesis of Dark Matter'' (680-47-532).

\end{acknowledgements}

\appendix
\section{Details of \Gadget simulations}
\label{sec:Gadget}

Here, we describe in more detail the setup of our numerical simulations. Parameter files for the simulations and scripts for generating initial conditions and analysing the outputs are publicly available \href{https://github.com/bradkav/BlackHolesDarkDress}{here} \cite{DarkDressCode}. Animations of selected simulations are also available \href{https://doi.org/10.6084/m9.figshare.6298397}{here} \cite{Animations}.

The simulations are performed using the \Gadget code \cite{Springel:2005mi} in a static (non-expanding) background, using only the tree-force algorithm to calculate gravitational forces without hydro-dynamical forces. We use the publicly available \textsc{pyGadgetIC} \cite{pyGadgetIC} as an interface to specify initial conditions and \textsc{pyGadgetReader} \cite{2014ascl.soft11001T} to read and analyse the \Gadget snapshots.

\begin{figure}[t]
\centering
\includegraphics[width=0.5\textwidth,]{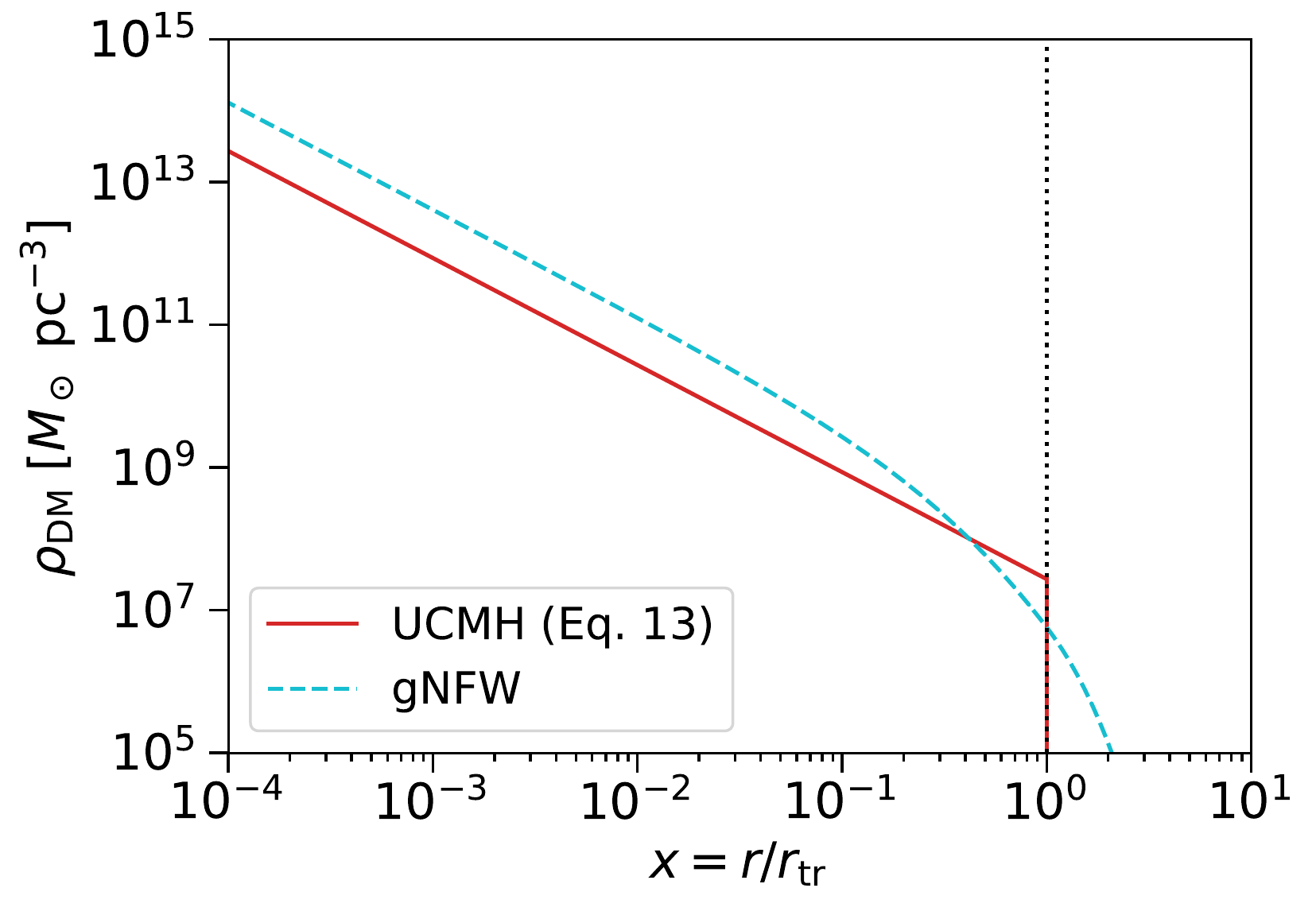}
\caption{\textbf{Dark Matter halo density profile.} The solid red line shows the density of DM, given by Eq.~\ref{eq:rhoDM}, inside the truncation radius $r_\mathrm{tr}$ for a $30 \,M_\odot$ PBH. The dashed blue line shows the approximate NFW-like profile used in our simulations, given by Eq.~\eqref{eq:gNFW}. The two profiles contain the same total mass and agree within a factor of 5 everywhere inside $R_\mathrm{tr}$.}
\label{fig:density_profile}
\end{figure}

As described in the main text, the DM halo should have a density profile $\rho \sim r^{-3/2}$ within the truncation radius $R_\mathrm{tr}$\cite{Lacki:2010zf}, flattening at larger radii. Such a density profile is not an equilibrium configuration in isolated conditions under Newtonian gravity. It is therefore unsuitable for simulating the evolution of the binary system over long timescales. In our \Gadget simulations, we approximate the DM density profile as a generalised NFW profile:
\begin{equation}
\label{eq:gNFW}
\rho(x = r/\Rtr) = 
\begin{cases}
\frac{\rho_0}{x^{3/2}(1+x)^{9/2}} & \text{for}\;x < 1\,,\\
a e^{-x/b}&  \text{for}\; x > 1\,.
\end{cases}
\end{equation}
This ensures that close to the PBH, the density profile rises as $ r^{-3/2}$, while at the truncation radius ($x = 1$), the density drops more rapidly ($\rho \sim r^{-6}$). At $x = 1$, we match onto an exponentially falling density profile, which ensures that the total halo mass is finite and the size of the halo is $\sim R_\mathrm{tr}$. We fix the normalisation $\rho_0$ by fixing the total halo mass at redshift $z$, given by Eq.~\eqref{eq:Mhalo}. We fix the constants $a$ and $b$ by requiring continuity of $\rho$ and $\partial \rho/\partial r$. In Fig.~\ref{fig:density_profile}, we illustrate the halo density profile at $z = z_\mathrm{eq}$ (red solid) and the density profile used in our simulations (blue dashed). 

We begin the simulation during the first in-fall of the PBHs, after a time-of-flight $t \sim 0.25\,t_\mathrm{orb}$--$0.41\,t_\mathrm{orb}$ from periapsis, where $t_\mathrm{orb}$ is the expected orbital time assuming point objects. More technically, we start the simulation at an eccentric anomaly of $u = 3\pi/4$ (i.e.~a quarter of the distance round the orbit from periapsis, when the velocities of the PBHs are anti-parallel). We have checked explicitly (using simulations of the full orbit) that tidal effects on the DM halo during this initial in-fall period are negligible.


For an orbit specified by the elements $(a_i,\, e_i)$, we require that the DM halos are resolved and stable down to radii smaller than the close-passage distance $r_\mathrm{min} = a_i(1-e_i)$. This means that we require that the softening length of the DM particles is at least a factor of a few smaller than the close passage distance. We also fix the number of DM particles to be 256 within a radius $r < 15\,r_\mathrm{soft}$ to ensure that the inner profile is sampled sufficiently. For small eccentricities, the close passage distance is larger, so only a coarse resolution is needed, while more eccentric orbits require a finer resolution. We therefore perform `low-resolution' simulations for $e > 0.995$ and `high-resolution' simulations for $e \leq 0.995$. 

\begin{table}[t]
\begin{ruledtabular}
\begin{tabular}{llll}
\texttt{ErrTolForceAcc} & \multicolumn{3}{l}{$10^{-5}$}\\
\texttt{ErrTolIntAccuracy} & \multicolumn{3}{l}{$10^{-3}$}\\
\texttt{MaxTimestep} [yr] & \multicolumn{3}{l}{$10^{-2}$}\\
$\ell_\mathrm{soft}$ (PBH) [pc] &  \multicolumn{3}{l}{$10^{-7}$}\\
\hline
$M_\mathrm{PBH} = $& $1 \,M_\odot$ & $30 \,M_\odot$ & $1000 \,M_\odot$ \\
\hline
$\ell_\mathrm{soft}$ (DM, low-res) [pc] & $2\times 10^{-6}$&$10^{-5}$ & $5\times 10^{-5}$\\
$\ell_\mathrm{soft}$ (DM, high-res) [pc] & $2\times 10^{-7}$ & $10^{-6}$ & $5\times 10^{-6}$\\
\end{tabular}
\end{ruledtabular}
\caption{\textbf{Summary of \textsc{Gadget-2} parameters.} The parameters \texttt{ErrTolForceAcc} and \texttt{ErrTolIntAccuracy} control the accuracy of force calculation and time integration respectively. We also specify the softening lengths $\ell_\mathrm{soft}$ of the simulations, as described in the text. Low-resolution simulations contain roughly $10^4$ DM particles per halo, while high-resolution simulations use a multi-mass scheme with roughly $4 \times 10^4$ DM particles in total per halo.}
\label{tab:Gadget}
\end{table}

For `low-resolution' simulations, we use $N_\mathrm{DM} \approx 10^4$ equal-mass DM particles per halo. If the DM psuedo-particles are too heavy the central BH undergoes stochastic motion \cite{Chatterjee:2002bg}. The number $N_\mathrm{DM}$ was chosen to guarantee the stability of the central BH particle within the halo, ensuring that the displacement of the BH from the centre of the halo does not exceed $r_\mathrm{min}$ over the relevant timescales\footnote{We note that specifying the maximum size of timesteps (see \texttt{MaxTimestep} in Tab.~\ref{tab:Gadget}) to be small also helps suppress the spurious stochastic motion of the central BH particle.}. The softening lengths used for the BH and DM particles for different set-ups are listed in Tab.~\ref{tab:Gadget}.

For `high-resolution' simulations, we use a multi-mass scheme \cite{Zemp:2007nt} in which the DM halo is composed of 4 different masses of pseudo-particle. The halo is divided into 4 shells, logarithmically spaced in radius. The inner-most shell has radius $r < 15\,r_\mathrm{soft}$ and the outer-most shell begins at radius $r > 0.1 \,R_\mathrm{tr}$. The inner-most shell is populated with the lightest particles, where again we use 256 particles to ensure a fine enough sampling. Particles initialised in the outer shells are more massive by a factor of 2 than those in the previous shell. This allows us to sample down to smaller radii using a factor of $\sim 7$ fewer pseudo-particles. We do not implement an orbit-refinement scheme, as suggested in Ref.~\cite{Zemp:2007nt}, but we confirm that the halo is sufficiently stable over the timescales of interest. The softening lengths for particles in the inner-most shell are given in Tab.~\ref{tab:Gadget}.

We note that in principle, it should be possible to resolve down to very small scales by increasing the number of DM particles per halo and decreasing the softening length. With sufficiently high resolution, it should then be possible to realistically simulate highly eccentric ($e > 0.9999$) binaries, relevant for mergers today. However, our current high-resolution simulations typically take in excess of 200 hours on 16 CPUs ($\sim 3000 \,\,\text{CPU-hours}$). Unfortunately, the runtime is not substantially improved by increasing the number of CPUs due to the highly clustered nature of the problem (most of the computational time is consumed by particles very close to the central PBH, of which there are only a small number). Furthermore, the dynamical timescale for particles close to the central black hole scales as $t_\mathrm{dyn} \sim r^{3/2}$ (for Keplerian motion), meaning that increasing the resolution becomes increasingly expensive at smaller and smaller radii. For a $30 \,M_\odot$ BH and DM particles at a radius $10^{-7}\,\mathrm{pc}$ (the value of $r_\mathrm{min}$ which gives mergers today), we have $t_\mathrm{dyn} \sim 10^{-4}\,\mathrm{yr}$ compared to the orbital time of the whole system $t_\mathrm{orb} \sim 10^{4} \,\mathrm{yr}$. Such simulations appear to be unfeasible with the setup we have used. Nonetheless, the simulations we have presented have sufficient resolution to capture the relevant physics and guide our analytic understanding.

\bibliography{refs}

\end{document}